\documentclass[prb,aps,10pt,twocolumn,amsmath,amssymb,superscriptaddress,longbibliography]{revtex4-2}
\usepackage[T1]{fontenc}
\usepackage[utf8]{inputenc}
\usepackage{geometry}
\usepackage{amsmath}
\usepackage{amsthm}
\usepackage{amssymb}
\usepackage{graphicx}
\usepackage{color}
\usepackage{physics}
\usepackage{comment} 
\usepackage{relsize}

\usepackage{hyperref}
\hypersetup{
  colorlinks   = true, 
  urlcolor     = black, 
  linkcolor    = blue, 
  citecolor    = red 
}

\usepackage[usenames,dvipsnames]{xcolor}


\usepackage[english]{babel}

\usepackage{hyperref}

\begin{document}

\title{BCS-BEC crossover in trapped one-dimensional Fermi–Hubbard chains: entanglement and correlation signatures from DMRG and effective-pairing theory}

\author{G. Diniz}
\affiliation{Charles University, Faculty of Mathematics and Physics, 121\,16 Prague, Czech Republic}
\author{I. M. Carvalho}
\affiliation{S\~ao Paulo State University (UNESP), Institute of Chemistry, 14800-090, Araraquara, S\~{a}o Paulo, Brazil}
\affiliation{The Abdus Salam International Centre for Theoretical Physics (ICTP), Trieste, Italy}
\author{M. Sanino}
\affiliation{S\~ao Paulo State University (UNESP), Institute of Chemistry, 14800-090, Araraquara, S\~{a}o Paulo, Brazil}
\affiliation{Okinawa Institute of Science and Technology, 1919-1 Tancha, Onna-son, Kunigami-gun, Okinawa 904-0495, Japan}
\author{F. Iemini}
\affiliation{Instituto de F\'{i}sica, Universidade Federal Fluminense, 24210-346 Niter\'{o}i, Brazil}
\author{V. V. Fran\c{c}a}
\affiliation{S\~ao Paulo State University (UNESP), Institute of Chemistry, 14800-090, Araraquara, S\~{a}o Paulo, Brazil}
\affiliation{The Abdus Salam International Centre for Theoretical Physics (ICTP), Trieste, Italy}

\date{\today}

\begin{abstract}
{Confined ultracold atoms in optical lattices provide a versatile platform for simulating lattice models of strongly correlated quantum systems, where pairing phenomena and superfluid phases can be explored under controlled conditions. While the crossover between the Bardeen–Cooper–Schrieffer (BCS) phase and the Bose–Einstein condensation (BEC) is well understood in homogeneous systems, spatial confinement breaks translational symmetry and reshapes correlation patterns, making the BCS-BEC identification in trapped geometries challenging and allowing unconventional phases to emerge with no direct analog in homogeneous systems. Here we present a characterization of the BCS–BEC crossover in harmonically confined one-dimensional Fermi–Hubbard chains. Our analysis combines Density Matrix Renormalization Group (DMRG) simulations and entanglement-based diagnostics with effective models describing the formation of tightly bound fermion pairs. This combined approach enables a detailed understanding of how the interplay between interactions and confinement reshapes the crossover, leading to insulating regions coexisting with persistent superfluid correlations. Within this framework, we further introduce conditioned correlation functions whose power-law decay allows a clear distinction between BCS-like and BEC-like regimes. The consistency between the effective descriptions and the numerical DMRG results yields a unified picture of the crossover in harmonically confined geometries. }

\end{abstract}

\maketitle

\section{Introduction}

The study of superconducting phenomena has long been a central topic in condensed matter physics, particularly following the discovery of high-temperature superconductivity \cite{RevModPhys.62.113,PhysRevLett.104.066406,PhysRevB.69.184501,Chin2006_,PhysRevResearch.2.023210, PhysRevB.105.184502}. In this context, the crossover between the Bardeen–Cooper–Schrieffer (BCS) regime of weakly bound Cooper pairs and the Bose–Einstein condensate (BEC) regime of tightly bound molecules in Fermi–Hubbard chains offers a valuable framework to investigate pairing mechanisms across interaction strengths and to explore the interplay between pairing and strong correlations \cite{Mizukami2023,PhysRevB.89.224508,PhysRevLett.92.130403,PhysRevB.60.3499}.

Homogeneous attractive Fermi–Hubbard chains with weak onsite interactions ($|U| \ll 2t$) are well described by BCS theory, where Cooper pairs have large correlation lengths \cite{Nozieres1985_,PhysRevB.59.7458}.  As $|U|$ increases, the system evolves smoothly from the BCS regime toward a state of tightly bound local pairs, characterizing the BCS–BEC crossover~\cite{Nozieres1985_,PhysRevB.59.7458,PhysRev.50.955,Vignolo2001_1DFermionsHarmonic}. This evolution is well understood in translationally invariant systems and typically takes place when the interaction strength becomes comparable to the bandwidth~\cite{Leggett1980_,10.1143/PTP.48.2171,Chen2005_BCSBEC,Randeria2014}. 
As a reference, for one-dimensional (1D) half-filled chains, where the bandwidth is $4t$, this gives the estimate $|U|\sim4t$~\cite{Leggett1980_}.

From an experimental perspective, ultracold atoms in optical lattices provide a highly tunable realization of the Hubbard model \cite{Roati2008_,Bakr2009_,PhysRevLett.124.010403,Hartke2023_,PhysRevLett.100.250403,PhysRevLett.116.033002,PhysRevLett.111.080501}, where two hyperfine states emulate spin-up and spin-down electrons, the lattice depth controls the hopping $t$, and Feshbach resonances tune the interaction $U$ \cite{Zwerger2003_MottHubbardColdAtoms,Brown2019_,PhysRevLett.125.113601}. On the one hand, this makes such systems ideal for investigating pairing and quantum phase transitions; on the other hand, the presence of a confining potential breaks translational symmetry and gives rise to inhomogeneous physics \cite{Greiner2002_SFtoMott,Gemelke2009_InSituMott,nair2024,nair2025}, including the suppression of long-range correlations and the modification of the momentum distribution, thus substantially altering the ground-state properties \cite{PhysRevB.82.014202,Guidini2016_BCSBEC_QCS}. As a consequence, standard probes of the BCS-BEC crossover --- based on pair correlation lengths, momentum distributions, or the chemical potential (Leggett criterion \cite{Leggett1980_}) ---  must be applied with caution, thereby motivating the development of new theoretical approaches \cite{PhysRevA.91.043612}.

In trapped one-dimensional systems, the loss of integrability generally prevents exact analytical solutions \cite{PhysRevLett.98.070402,PhysRevB.82.014202}. Few-particle simulations based on exact diagonalization or Quantum Monte Carlo have provided valuable insights into the evolution of bound states and energetics across the crossover in confined geometries \cite{PhysRevLett.92.160401,Blume2012_FewBodyUltracold,PhysRevA.76.053613,PhysRevLett.105.110401}. Mean-field variational methods have also been extended to include spatial inhomogeneity \cite{PhysRevA.91.043612}. Local-density approximations (LDA) based on homogeneous Bethe–Ansatz solutions often capture qualitative features of density profiles, entanglement, and global observables \cite{Canella2019_,Pauletti2024_,PhysRevA.92.013614,Capelle_}. {More robust numerical techniques, such as the Density Matrix Renormalization Group (DMRG) for 1D systems and Dynamical Mean-Field Theory for higher-dimensional lattices, have revealed new confinement-induced behaviors \cite{PhysRevB.76.220508,PhysRevB.72.235118,PhysRevB.82.014202,PhysRevA.89.053604,PhysRevB.79.214518,sanino2024entanglement}}. Yet, despite these advances, establishing robust and physically intuitive criteria to identify and classify the BCS–BEC crossover in strongly inhomogeneous many-body systems remains an open challenge.

Here we investigate the BCS–BEC crossover in the 1D harmonically confined Fermi–Hubbard chain at zero temperature, combining DMRG and effective pairing models across the full range of interaction strengths, from weak-coupling
to strong-coupling regime. We derive a confined tight-binding model describing hard-core bosons that captures the system’s behavior in the BEC regime and provides a microscopic explanation {\it i)} for the effective enhancement of the trapping potential at large $|U|$ \cite{PhysRevA.89.053604}, and {\it ii)}  for the robustness of the superfluid topology observed in trapped chains \cite{sanino2024entanglement}. Mean-field analysis reveals an unusual BCS-like behavior in which pairs avoid the trap center and localize at the interface between empty sites and the central region, accounting for the superfluid wings observed in Ref.~\cite{PhysRevB.82.014202} at small and intermediate $|U|$. For strong interactions, a similar phenomenology arises from effective repulsion between BEC-like quasiparticles \cite{RevModPhys.62.113}, consistent with previous observations \cite{sanino2024entanglement}. 

\begin{figure}[h!]
\begin{centering}
\includegraphics[scale=0.57]{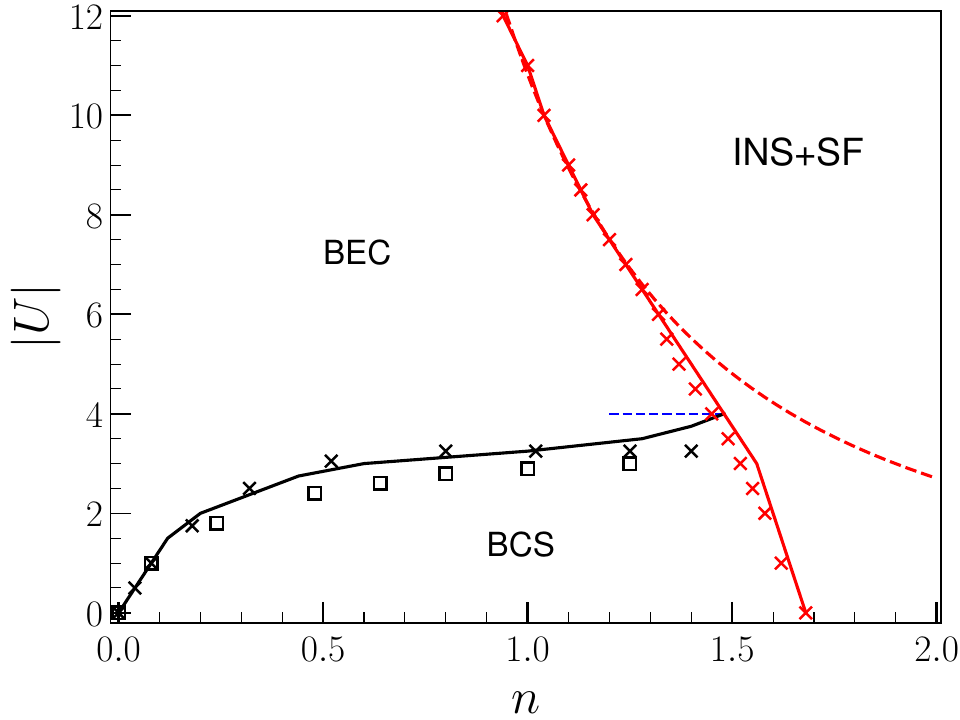}
\par\end{centering}
\caption{ Phase diagram predicted by the effective models as function of the interaction strength $|U|$ and average density $n$. Solid lines indicate the boundaries between BCS- and BEC-dominated regimes (black), using the criterion $r_{\mathrm{pair}} \le 1$ from Eq.~\eqref{pair_size_estimative}; and the insulator (INS) and the superfluid (SF) phase (red), adopting the criterion $\max_j(n_j) \ge 1.99$. For comparison we also include DMRG results: $\times$ symbols (entanglement-based method) and square symbols (correlator-based method). The red dashed line represents Eq.~\eqref{BEC-INS-PRED}, showing good agreement for $|U| \gtrsim 6t$; while the blue dashed line indicates the interaction value $|U|=4t$, above which no BCS regime is observed. All data are for fixed $k = 0.0005$ and $L = 100$ sites.
\label{Phase_Diagram1}}
\end{figure}

Our results show qualitative agreement between DMRG and effective models, which we use to propose new strategies for identifying the BCS–BEC crossover in confined systems, {based on the dominance of BCS- or BEC-like pairing behavior.} 
These include a four-particle correlation function, previously unexplored in this context, that captures the BCS–BEC crossover through distinct power-law decays in each regime, as well as half-chain entanglement measures that identify both the crossover and the superfluid–insulator transition, highlighting the relevance of entanglement-based probes for many-body phase transitions. {Figure~\ref{Phase_Diagram1} summarizes the crossover and phase transitions in the harmonically confined chain. At low densities, any finite attractive interaction leads to tightly bound BEC pairs within the superfluid (SF) phase. As the density increases, screening by unpaired electrons drives a crossover to longer-range BCS pairing. Stronger interactions stabilize BEC pairing up to higher densities, eventually suppressing the BCS regime. At higher enough densities, the trap center becomes insulating (INS), while superfluid states persist at the edges.}

Our work is organized as follows. Section II introduces the models and methods. Section III develops the effective-model analysis, providing physical intuition and a qualitative phase diagram. Section IV presents the DMRG results, which confirm and refine this picture through a fully many-body treatment. Section V concludes with a summary of our findings.

\section{Models and Methods}\label{sec_model}

We consider the 1D Fermi–Hubbard model with harmonic confinement, 
\begin{align}\label{Eq. 1}
    H = &- t \sum_{\sigma,j} ( c_{j,\sigma}^\dagger c_{j+1,\sigma} + \mathrm{h.c.}) + U\sum_{j} n_{j,\uparrow} n_{j,\downarrow} \nonumber \\ 
    &+ t\sum_{\sigma, j}  k \left(j-j_0\right)^2  c_{j,\sigma}^\dagger c_{j,\sigma},
\end{align}
where $U<0$ is the on-site attractive interaction, the index $1 \le j \le L$ labels the sites along the chain, where $L$ is the total number of sites, and $tk(j-j_0)^2$ represents the confinement potential centered at $j_0=(L+1)/2$. The operator $c_{j\sigma}^\dagger$ ($c_{j\sigma}$) creates (annihilates) an electron with spin $\sigma = \uparrow, \downarrow$ at site $j$, and $n_{j\sigma}$ is the corresponding number operator. Throughout this work, we set $t = 1$, which defines the unit of energy.

 The ground-state properties of this Hamiltonian are numerically accessed via DMRG calculations in its matrix product state (MPS) formulation~\cite{dmrg, SCHOLLWOCK_MPS}. In this approach, the accuracy is controlled by the maximum bond dimension $\chi_{\mathrm{max}}$, which we set initially to $1500$ and progressively increase during the sweeps. Simulations are performed for non-magnetic (spin-balanced) systems, with the total number of electrons $N = nL$ conserved and equally distributed between spins ($N_\uparrow=N_\downarrow$) throughout the optimization. The initial MPS places doubly occupied sites at the chain center, consistent with the expected distribution under both attractive $U$ and the parabolic trap, ensuring fast energy convergence ($\sim10^{-7}$) and stable central-bond entropy ($\mathcal{O}(10^{-5})$). A warm-up stage of ten sweeps with small $\chi$ values precedes the full optimization, ensuring a smooth buildup of entanglement. All DMRG calculations are performed with open boundary conditions using the \textsc{ITensor} library~\cite{itensor}.

To understand the pairing mechanism in confined scenarios, we also analyze simpler effective models, as described in the following subsections.

\begin{figure*}[!ht]
\begin{centering}
\includegraphics[scale=0.6]{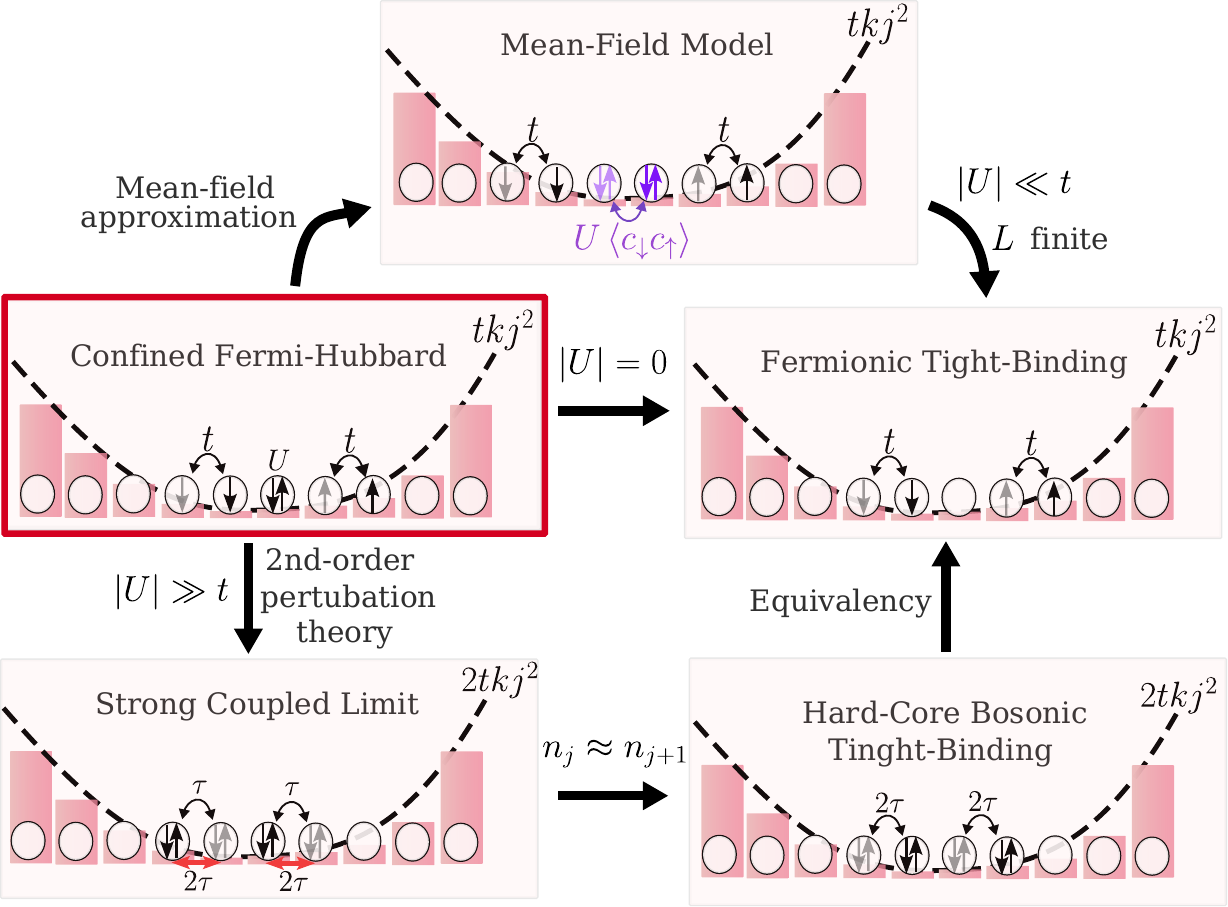}
\par\end{centering}
\caption{Schematic representation of the effective models derived from the \textbf{confined Fermi-Hubbard} model in Eq.~\eqref{Eq. 1}. \textbf{Strongly coupled limit:} For $|U| \gg t$, perturbation theory yields an effective description of the BEC regime (Eq.~\eqref{EFF_SCL}) with confinement strength $2k$, hopping $\tau = 2t^2/|U|$, and nearest-neighbor repulsion $2\tau$. \textbf{Hard-core bosonic tight-binding:} When the density varies smoothly, this model further reduces to a confined tight-binding chain with hopping $2\tau$ (Eq.~\eqref{EFF_SCL_TTB}).  Through the
Girardeau mapping \cite{Girardeau1960_}, together with the renormalized confinement strength $\tilde k = |U|k/2t$, the system can equivalently be described as a confined spinless \textbf{fermionic tight-binding} chain. For $|U| \ll t$ and finite system size $L$, no coherent pairs are formed, so the system behaves as a confined tight-binding chain. \textbf{Mean-field model:} For finite $|U|$ up to intermediate values, mean-field theory provides an appropriate description, allowing both single-particle hopping and pair hopping via the term $U\langle c_{\downarrow} c_{\uparrow} \rangle$ (Eq.~\eqref{EFF_WCL}).
\label{Schematics}}
\end{figure*}

\subsection{Effective model for $|U| \gg t$}

For the strongly coupled regime, the system approaches the atomic limit, where the electron tunneling between sites is suppressed ($t/|U| \rightarrow  0$), intersite correlations become weak, and the ground state is dominated by tightly bound on-site electron pairs. Nevertheless, a residual coupling between neighboring sites is  induced by virtual hopping processes of order $\tau\approx 2t^{2}/|U|$, arising from second-order tunneling events. In this regime, individual electron transport is suppressed, and charge dynamics occur through the coherent motion of bound electron pairs along the chain (see details in Section~2 of the Supplementary Material (SM)). These pairs, described by the hard-core bosonic creation operator $\phi_j^\dagger = c_{j\uparrow}^\dagger c_{j\downarrow}^\dagger$\footnote{This quasiparticle behaves as a spin-zero hard-core boson. Operators on different sites commute, while double occupancy of the same site is forbidden due to the underlying fermionic structure of the pair.}, constitute the relevant quasiparticles of the system.

In this regime, second-order perturbation theory in $t/|U|$ leads to the following effective Hamiltonian \cite{RevModPhys.62.113}:
\begin{align}\label{EFF_SCL}
   \tilde H &\equiv \sum_j \left[ -|U|(1+4t^2/|U|^2) + 2tk(j-j_0)^2 \right] \phi_{j}^\dagger  \phi_{j} \nonumber \\  &-\frac{2t^2}{|U|} \sum_{j} (\phi_{j}^\dagger \phi_{j+1} +\mathrm{h.~c.})+ \frac{4t^2}{|U|} \sum_{j}\phi_{j}^\dagger \phi_{j} \phi_{j+1}^\dagger \phi_{j+1},
\end{align}
describing a confined tight-binding model of hard-core bosonic quasiparticles with weak nearest-neighbor repulsion. Notice that for $k=0$ and  unit filling ($n=1$), the model becomes analogous to the  Heisenberg model obtained in the repulsive ($U>0$)  case \cite{RevModPhys.62.113}, up to corrections of order $t^{4}/|U|^{3}$.

\subsection{Mean-field approximation for $|U|\lesssim 2t$ }

In the opposite limit, where $|U|$ is comparable to or smaller than the hopping amplitude $t$, the mean-field (MF) approximation has proven to be useful for capturing qualitative features~\cite{BCS_paper,Nozieres1985_,PhysRevA.86.023619}. Within this approximation, the interaction term is decoupled, allowing the Hamiltonian to be written in the following MF form:
\begin{align}\label{EFF_WCL}
    H_{MF} \equiv \sum_{j \sigma}& e_j n_{j\sigma}  -t \sum_{j\sigma} \left( c_{j\sigma}^\dagger c_{j+1 \sigma} +\mathrm{h.~c.} \right)  \nonumber \\ -&U \sum_j  \left( \langle c_{j\downarrow}  c_{j\uparrow} \rangle\, c^\dagger_{j\uparrow} c^\dagger_{j\downarrow} +\mathrm{h.~c.} \right), 
\end{align}
where $e_j = -\mu + tk\left( j - j_0\right)^2 + U \langle n_j \rangle$. This Hamiltonian is solved self-consistently through iterative diagonalization and parameter updates until convergence is achieved. In particular, the chemical potential 
$\mu$ is adjusted at each step to ensure that the total particle number remains close to the target value (see SM Section~4 for details) \cite{PhysRevC.96.014305,Sheikh2021}. 

\section{Effective-Model Results}

To build physical intuition and clarify the pairing mechanisms, we first analyze the effective models, which also provide a useful framework for identifying crossover signatures in the numerical results. A schematic summary of these models, together with their assumptions and regimes of validity, is shown in Figure ~\ref{Schematics}.

\subsection{Metal–insulator transition for $U=0$}

\begin{figure}[!ht]
\begin{centering}
\includegraphics[scale=0.57]{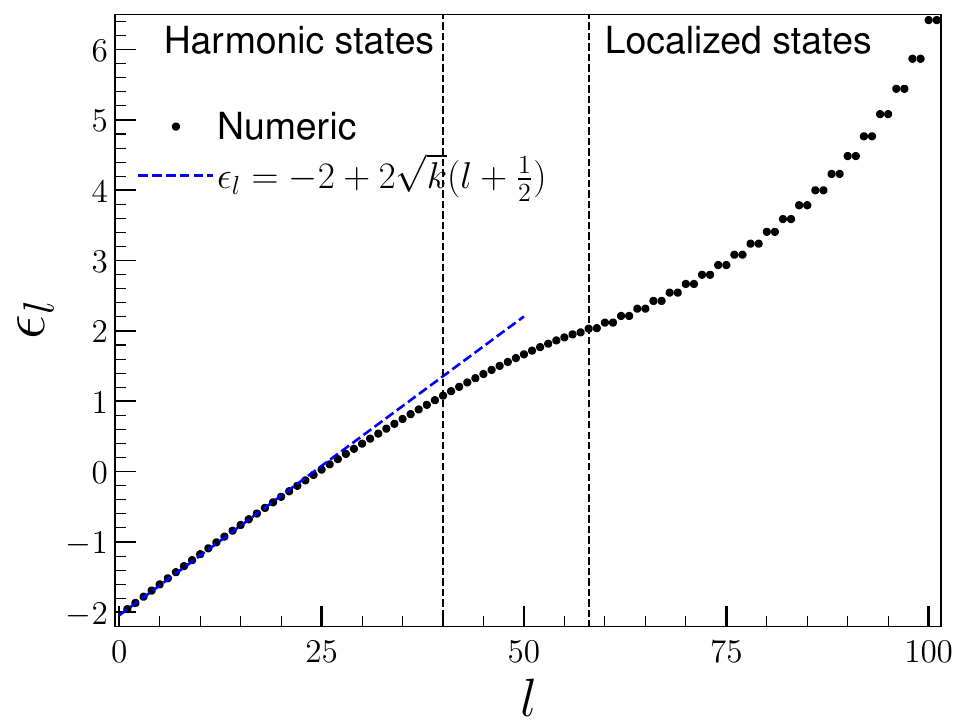}
\par\end{centering}
\caption{Spectrum ${\epsilon_l}$ of the confined tight-binding model with $k=0.002$ and $L=100$, where $l$ indexes eigenstates by increasing energy. The blue dashed line shows a harmonic fit, while the black dashed lines mark the boundary between harmonic and localized states. Notice that the localized states are degenerate due to the system's even symmetry.
\label{TTB_Spectrum}}
\end{figure}

We start by examining the non-interacting limit ($U=0$), as it provides a useful reference point for understanding both the $U \ll t$ and the $|U| \gg t$ regimes, as shown in Fig. \ref{Schematics}. In this case, the Hamiltonian~\eqref{Eq. 1} reduces to a simple confined tight-binding model. As shown in section 5 of the SM, at low densities the chain behaves approximately as a quantum harmonic oscillator, with $t$ playing the role of kinetic energy. The low-energy states then follow a linear dispersion $\epsilon_l \approx -2t + 2t\sqrt{k}(l + 0.5)$, with $l$ labeling each state, as illustrated in Figure ~\ref{TTB_Spectrum}. 

This analogy, however, breaks down for energy levels $\epsilon_l \sim 2t$: as higher-energy harmonic states become occupied, their wavefunctions develop progressively faster oscillations. Once these oscillations reach a wavelength comparable to the lattice spacing, the corresponding states no longer perceive an effectively continuous potential, but rather the underlying discrete translational symmetry of the chain, causing the wavefunctions to become localized (see SM section 5). This breakdown defines the positions $j^*_\pm \approx \frac{L}{2} \pm \frac{1.3}{\sqrt{k}}$, at which the wavefunctions start to become localized. Since the localized states do not affect the central density, they become accessible only after the region within $\pm 1.3/\sqrt{k}$ of the center is fully occupied with two particles per site. As a consequence, if $N = nL \ge 4 \times 1.3/\sqrt{k}$, that is, if the density exceeds a critical value given by
\begin{align}\label{Metal-INS-PRED}
    n_c \equiv {5.2} \sqrt{\frac{1}{kL^2}}
\end{align}
the chain enters a charge-insulating regime, since charge carriers can no longer propagate freely through the central region, where all accessible low-energy states are already occupied.

\subsection{Superfluid–insulator transition in BEC regime} 

Now, let us consider the $|U|\gg t$ limit, where the BEC pairs behave nearly as non-interacting quasiparticles \cite{Vignolo2001_1DFermionsHarmonic}. While Eq.~\eqref{EFF_SCL} describes the ground-state properties, its interaction term makes a direct analysis challenging. We therefore introduce a simpler model (see SM, section 3), derived under the assumption of a smooth density profile, with the Hamiltonian given by:
\begin{align}\label{EFF_SCL_TTB}
    H_{\mathrm{BEC}} =\frac{4t^2}{|U|} \left(\sum_j \tilde k (j-j_0)^2 \phi_j^\dagger \phi_j -\sum_j (\phi_j^\dagger \phi_{j+1} + \mathrm{h.c.}) \right).
\end{align}
This model provides a simple confined tight-binding description for the BEC quasiparticles, with effective hopping $2\tau \approx 4t^{2}/|U|$ (arising from virtual processes), confinement $\tilde{k} = k|U|/2t$ and average density $\tilde{n} = 0.5n$, yielding a quadratic Hamiltonian that is straightforward to diagonalize.

\begin{figure}[tbh!]
\begin{centering}
\includegraphics[scale=0.57]{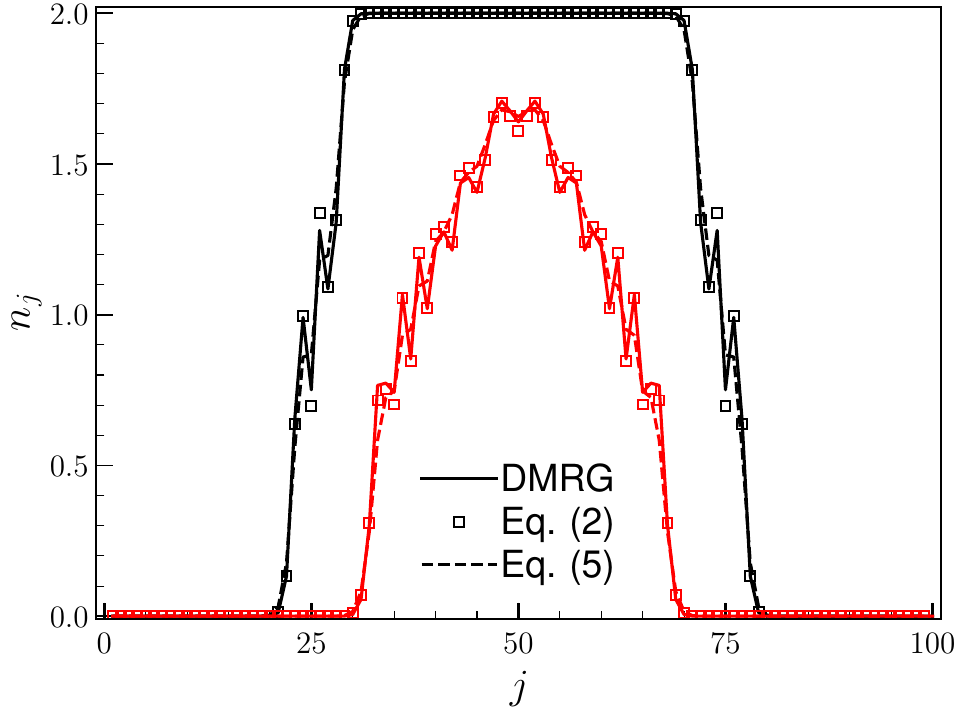}
\par\end{centering}
\caption{Density profile $n_j = \langle \sum_\sigma c_{j\sigma}^\dagger c_{j\sigma}\rangle$ along the chain for  $n=1$ (black curves) and $n=0.46$ (red curves) at fixed $U=-10.0t$ and $k=0.002$. The results were obtained via DMRG (solid lines), via the effective mean-field model in Eq.~\eqref{EFF_SCL} (square symbols), and via the effective tight-binding model in Eq. \eqref{EFF_SCL_TTB} (dashed lines).
\label{BEC_Density}}
\end{figure}

Figure ~\ref{BEC_Density} shows the density profile along the chain obtained from the effective models and from DMRG in the BEC regime for two average densities. The results from Eq. \eqref{EFF_SCL} show a maximum percentage error of $5\%$, reflecting the approximation inherent to the mean-field approach; an exact solution would be expected to yield a smaller error. In comparison, the effective tight-binding model in Eq.~\eqref{EFF_SCL_TTB} exhibits a maximum error of $15\%$. Both errors peak around the interface between empty sites and the trap center, hereafter referred to as \emph{edges}, where fast density oscillations are observed. Apart from the deviations in the edges, the good agreement with DMRG indicates that the renormalization of the confinement strength with $|U|$, as given by Eq.~\eqref{EFF_SCL_TTB}, qualitatively accounts for the increase in the density at the center of the chain observed in Refs.~\cite{PhysRevA.89.053604, PhysRevA.91.043612}.

Notice that, while the density oscillations at the edges are pronounced and comparable between the DMRG results and those from in Eq.~\eqref{EFF_SCL}, they are significantly smoother in the results obtained from Eq.~\eqref{EFF_SCL_TTB}. Since the only difference between Eqs.~\eqref{EFF_SCL} and \eqref{EFF_SCL_TTB} is the presence of a nearest-neighbor repulsion between BEC pairs, these oscillations can be attributed to this effective interaction. In homogeneous systems, translational symmetry ensures that the repulsion from left and right neighbors cancels out, rendering this term ineffective. In contrast, the confining potential breaks this balance, leading to a clear manifestation of the repulsive interaction. Thus this effective BEC-pairs repulsion accounts for the superfluid states that persist in the edges when $|U| \gg t$ ~\cite{sanino2024entanglement}. Although they resemble charge-density-wave–like modulations \cite{PhysRevB.105.115116,PhysRevLett.62.1407}, the oscillations always occur with quasi-momentum $q\approx\pi$ and do not depend on the filling, as evidenced in Fig.~\ref{BEC_Density}.

\begin{figure}[tbh!]
\begin{centering}
\includegraphics[scale=0.55]{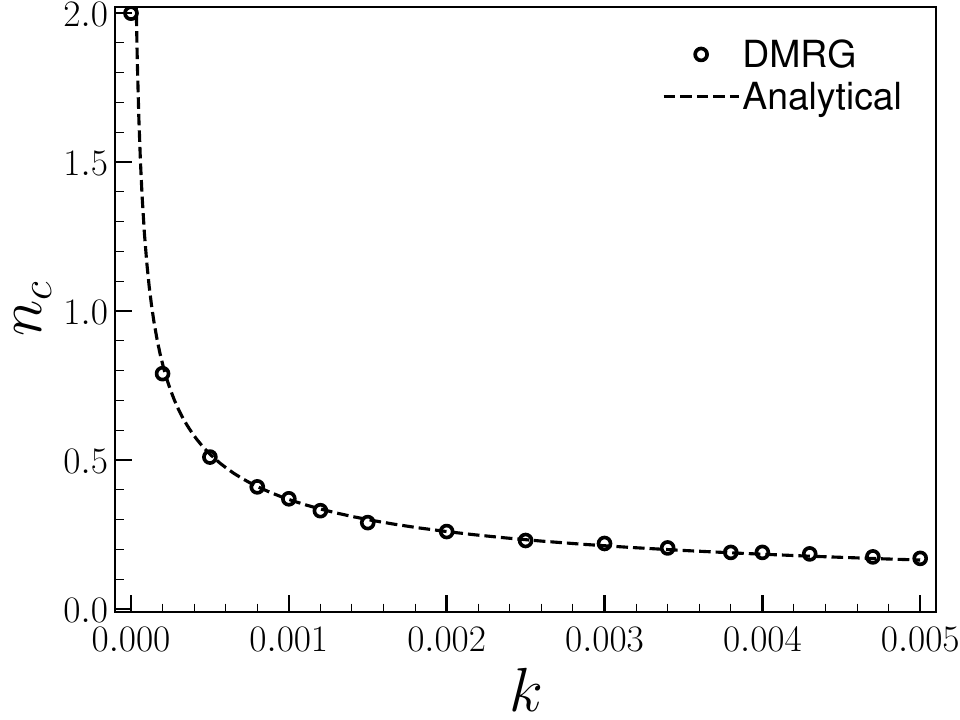}
\par\end{centering}
\caption{Critical densities $n_c$ for the superfluid–insulator transition obtained from DMRG calculations (circular dots) and from Eq.~\eqref{BEC-INS-PRED} (dashed line), for $U=-10$ and $L=200$. 
\label{n_c}}
\end{figure}

Now, taking advantage of the fact that Eq.~\eqref{EFF_SCL_TTB} reproduces the overall shape of the density profile in Fig.~\ref{BEC_Density}, one can apply the Girardeau mapping \cite{Girardeau1960_}, which establishes the equivalence between hard-core bosons and spinless fermions in 1D, and thereby use the prediction of Eq.~\eqref{Metal-INS-PRED} to estimate the superfluid–insulator transition. Therefore, when the density exceeds a critical value given by
\begin{align}\label{BEC-INS-PRED}
    n_c \equiv {5.2} \sqrt{\frac{2t}{|U|kL^2}},
\end{align}
the chain will behave as a charge insulator. Figure~\ref{n_c} compares the critical densities obtained from DMRG calculations --- via the charge gap as in Ref.~\cite{sanino2024entanglement} --- with those predicted by Eq.~\eqref{BEC-INS-PRED}, showing excellent agreement. Therefore, despite its simplicity, Eq.~\eqref{BEC-INS-PRED} is able to accurately predict the transition from superfluid in BEC regime to the insulating phase.

\subsection{Mean-field approach for the BCS-BEC crossover}

For given values of $U$, $n$, and $k$, the self-consistently diagonalization of the mean-field Hamiltonian~\eqref{EFF_WCL} leads to
\begin{align}\label{MFHD}
    H_{MF} = \sum_l \left[ \xi_l \gamma^\dagger_{l,+} \gamma_{l,+} - \xi_l \gamma^\dagger_{l,-} \gamma_{l,-} \right], 
\end{align}
where $\gamma^\dagger_{l,\pm}$ creates a Bogoliubov quasiparticle (BQP) with energy $\pm \xi_l$. These operators are linear combinations of a spin-$\uparrow$ electron and a spin-$\downarrow$ hole:
\begin{align}
    \gamma_{l,-}^\dagger  = \sum_j \left[ \alpha_{l,j} c_{j,\uparrow}^\dagger + \beta_{l,j} c_{j,\downarrow} \right],
\end{align}
\begin{align}
    \gamma_{l,+}^\dagger  = \sum_j \left[ -\beta_{l,j} c_{j,\uparrow}^\dagger + \alpha_{l,j} c_{j,\downarrow} \right].
\end{align}
Here, $\alpha_{l,j}$ and $\beta_{l,j}$ are the Bogoliubov amplitudes, satisfying $ \sum_j |\alpha_{l,j}|^2 + |\beta_{l,j}|^2 = 1$.

\begin{figure}[tbh!]
\begin{centering}
\includegraphics[scale=0.57]{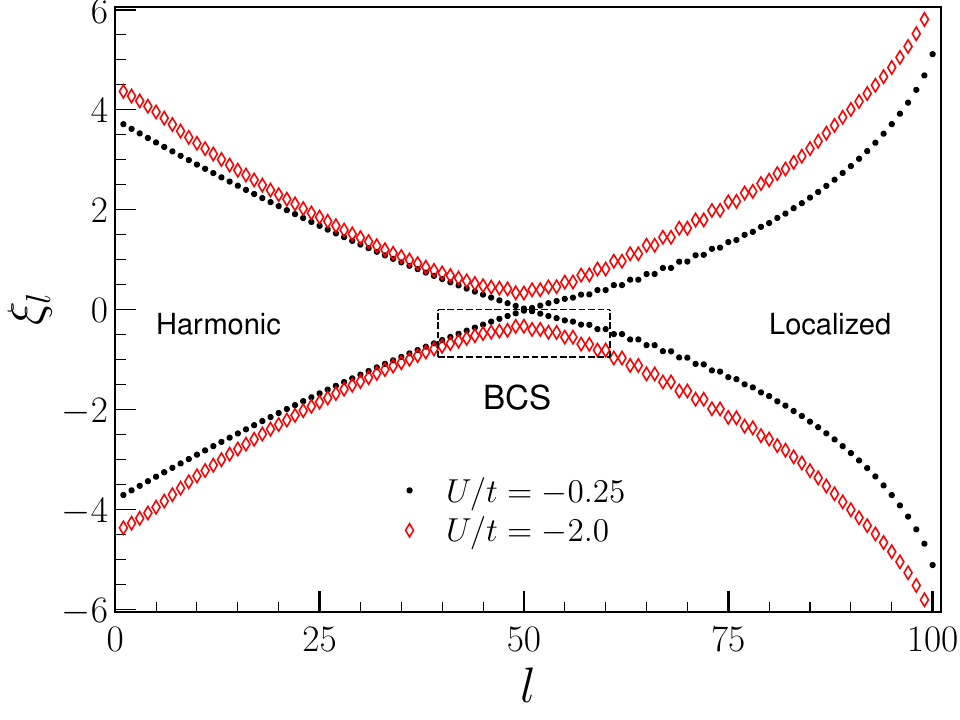}
\par\end{centering}
\caption{Mean-field quasiparticle spectrum for $U = -0.25t$ (black) and $U = -2t$ (red), with fixed $n=1$ and $k=0.002$. The dashed lines delimitates the region where pairing is concentrated for $U=-2t$. The left side shows harmonic-like states, while the right contains localized states, as also seen in the confined tight-binding chain in Fig.~\ref{TTB_Spectrum}.
\label{MF_Spectrum}}
\end{figure}

Thus, to understand how the system evolves from the BCS to the BEC regime, we examine in Figure ~\ref{MF_Spectrum} the self-consistent mean-field spectrum as $|U|$ increases for $k = 0.002$ at half filling. For $U=-0.25t$, the mean-field spectrum closely resembles the $U=0$ case, since the mixing between spin-$\uparrow$ particles and spin-$\downarrow$ holes is minimal. As $|U|$ increases, pairing is enhanced and a superconducting gap becomes visible. Moreover, pairing is no longer restricted to states near the gap, but extends to states farther away from it. Nevertheless, near the gap the quasiparticle states exhibit a strong mixing of spin-$\uparrow$ particle and spin-$\downarrow$ hole components, so that  paired levels with energies $\pm\xi_{l}$ encode the Cooper-pair structure characteristic of BCS theory.

\begin{figure}[t!]
\begin{centering}
\includegraphics[scale=0.57]{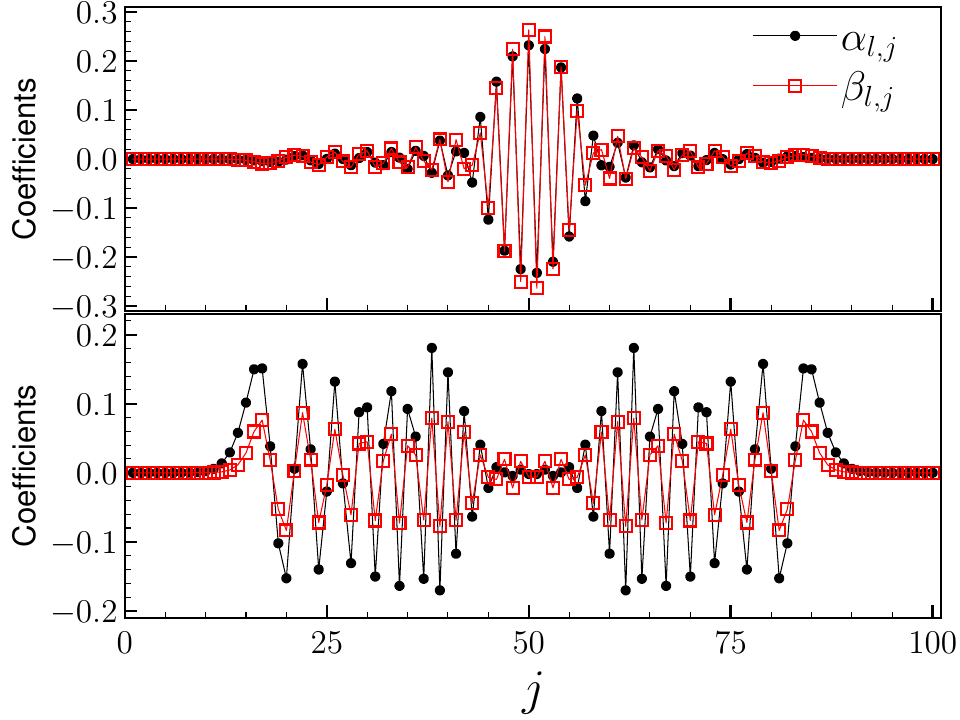}
\par\end{centering}
\caption{Bogoliubov coefficients along the chain for spin-$\uparrow$ particles  $\alpha_{l,j}$ (black circles) and spin-$\downarrow$ holes $\beta_{l,j}$ (red squares), for $U = -2t$, $k = 0.002$, $L=100$, and $n = 1$. The top panel shows the BQP wavefunction immediately below the gap ($l=50$), while the bottom panel shows a state near the localized-state region ($l=55$).
\label{MF_WF_k=002}}
\end{figure}

Figure~\ref{MF_WF_k=002} shows the BQP states near the gap and near the localized-state region \footnote{Note that $\alpha$ and $\beta$ show similar oscillations, leading to related momentum distributions. Since a hole with momentum $\vec{q}$ is equivalent to a particle with momentum $-\vec{q}$, the Bogoliubov quasiparticle acting on the effective vacuum encodes a BCS-like pair structure.}. The former behaves as a harmonically confined BCS pair, whereas the latter corresponds to a BCS-like pair that avoids the center of the chain. These unusual BCS pairs are associated with the oscillations observed in the density profile for small and intermediate values of $|U|$ (see SM, section 4).

Finally, to classify whether the system lies predominantly in the BEC or in the BCS regime within the mean-field formalism, it is useful to consider the average relative distance between the spin-$\uparrow$ and spin-$\downarrow$ components of a Cooper pair ~\cite{Strinati2018_BCSBEC_review,PhysRevB.65.014501}, defined as
\begin{align}\label{pair_size_estimative}
    r_{\mathrm{pair}}^2 = \dfrac{\mathlarger{\sum}_{r} r^2 \left| \mathlarger\sum\limits_{l / \xi_{l} < 0}  \mathlarger\sum\limits_{j} \alpha_{l,j} \beta_{l,j+r}\right|^2}{\mathlarger\sum_r \left|  \mathlarger\sum\limits_{l / \xi_{l} < 0}  \mathlarger\sum\limits_{j} \alpha_{l,j} \beta_{l,j+r}\right|^2}.
\end{align}
This quantity provides the average separation between spin-$\uparrow$ and -$\downarrow$ particles along the chain. Thus, for $r_{\mathrm{pair}} \gg 1$ the system is in the BCS regime, whereas for $r_{\mathrm{pair}} \le 1$, it predominantly lies in the BEC regime (see SM, section ~4).

\subsection{Phase diagram from effective models}

Figure~\ref{Phase_Diagram1} shown at introduction summarizes the predictions of the effective models as a function of the interaction $|U|$ and the density $n$. At very low densities, particles concentrate in the trap center, each spin-$\uparrow$ particle spatially overlaping with a spin-$\downarrow$ particle, resulting in a superfluid phase  characterized by tightly bound BEC pairs, with a coherence energy of order $\sim |U| + 2t\sqrt{k}$. This BEC behavior also appears in homogeneous systems \cite{Nozieres1985_}; however, confinement extends its stability to higher densities. In both cases, increasing $|U|$ further stabilizes the BEC regime.

As $n$ increases, for $0 < |U| \lesssim 4t$, the system remains in the superfluid phase but develops longer-range momentum-correlated pairs corresponding to extended Cooper pairs as described in BCS theory. Due to confinement, however, these pairs are not pure plane waves with well-defined momentum; instead, they exhibit a broadened momentum distribution (see Fig.~\ref{MF_WF_k=002})\footnote{The resulting state can therefore be interpreted as a superposition of different Cooper-pair configurations.}. As the density increases further, $n \ge n_c$, as indicated by the red curves, localized states start to form at the trap center, leading to an insulating (INS) behavior in this region, while superfluid (SF) states persist at the edges of the central insulating region \cite{PhysRevB.82.014202,sanino2024entanglement}. These edge states are robust and persist for all $n<2$, clearly distinguishing this situation from a conventional insulating phase and resulting in a composite INS+SF phase.


\section{DMRG RESULTS}\label{sec_results}

Next, we employ DMRG calculations to extend the analyses and validate the physical picture of Fig.~\ref{Phase_Diagram1}, thereby obtaining a more reliable description.

\subsection{BCS-BEC crossover from correlation functions}\label{correlator_method}

A standard way to probe pairing in DMRG is via correlation functions. However, in confined systems, conventional correlators, such as density–density or pair-pair correlations, do not clearly distinguish between the BEC and BCS regimes (see SM, section 6). To overcome this, we exploit the expected behavior in each regime ---tightly bound on-site pairs in the BEC limit, while in the BCS regime spin-$\uparrow$ fermions move more independently of the spin-$\downarrow$ background --- and employ a correlation function that directly probes the local binding between opposite spins:
\begin{align}\label{Correlator2}
    \mathcal{P}_{j}(r) = \langle c^\dagger_{j+r,\uparrow} c_{j,\uparrow} n_{j,\downarrow}\rangle.
\end{align}
This shows how the propagation of spin-$\uparrow$ fermions from site $j$ to $j+r$ is affected by the presence of spin-$\downarrow$ fermions at site $j$. Within the BEC regime one expects $\mathcal{P}_{j}(r)$  to decrease faster with $r$ than in the BCS regime. 

\begin{figure}[th!]
\begin{centering}
\includegraphics[scale=1.00]{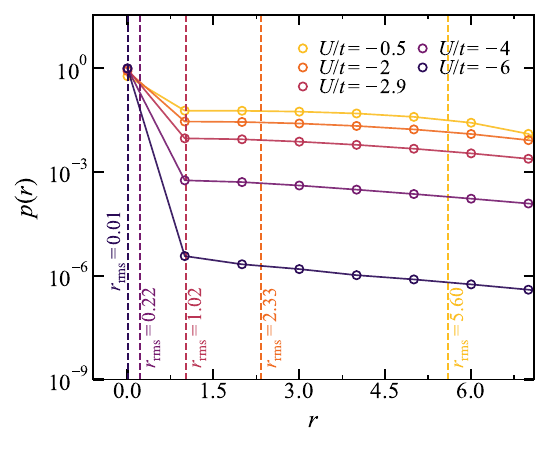}
\par\end{centering}
\caption{Distribution $p(r) = \dfrac{|\mathcal{P}_{L/2}(r)|}{\sum_{r'} |\mathcal{P}_{L/2}(r')|}$ for $r \ge 0$, obtained from Eq.~\eqref{Correlator2} with fixed $j=L/2$ (center of the chain, minimizing boundary effects). Results are shown for $k = 5 \times 10^{-4}$, $n = 1$, and $L = 100$. {Several values of $|U|$ are shown across the full crossover region to illustrate the behavior of the distribution beyond the limiting regimes (colors indicated in the legend)}.  The vertical dashed lines mark $r_{\mathrm{rms}} = \sqrt{\sum_{r} r^{2} p(r)}$, which provides a measure of the spatial extent of the pair distribution and serves as an estimate of the pair size. 
\label{Fig:Correlator1}}
\end{figure}

As shown in Figure \ref{Fig:Correlator1}, this is precisely the case: for $U=-0.5t$ the correlation remains significant at large $|r|$, while in the BEC regime ($U=-4t$) electrons remain correlated only for short distances. We find that the root-mean-square distance, which provides a measure of $p(r)$ spatial extent, decreases from $r_{\mathrm{rms}}=5.60$ at $U=-0.5t$ (BCS-like regime) to values smaller than unity, $r_{\mathrm{rms}}=0.22$ at $U=-4t$ (BEC-like regime). Thus $p(r)$ becomes increasingly sharp as $|U|$ increases, and therefore, $r_{\mathrm{rms}}$ directly reflects the progressive shrinking of the pair size across the BCS–BEC crossover.

\begin{figure}[th!]
\begin{centering}
\includegraphics[scale=0.79]{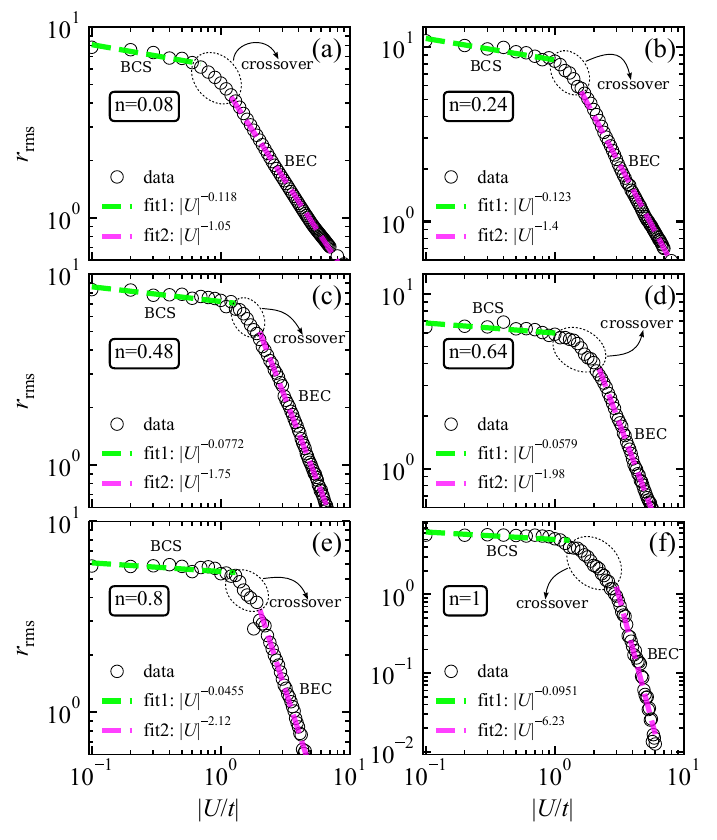}
\par\end{centering}
\caption{Root-mean-square distance $r_{\mathrm{rms}}$ of $p(r)$ as a function of $|U|/t$ on a log–log scale for several average densities $n$. Numerical results are shown as black circles, with densities indicated in each panel: $n=0.08$ (a), $n=0.24$ (b), $n=0.48$ (c), $n=0.64$ (d), $n=0.80$ (e), and $n=1.00$ (f). The green line shows a power-law fit $r_{\mathrm{rms}} \sim r_1 |U/t|^{-\nu_1}$ for weak interactions ($|U|/t \lesssim 1$), characterized by a small exponent, consistent with BCS-like regime). The purple line shows a second power-law fit $r_{\mathrm{rms}} \sim r_2 |U/t|^{-\nu_2}$ for stronger interactions ($|U|/t \gtrsim 2$), with a larger exponent, consistent with the BEC-like regime).
\label{Fig:Correlator2}}
\end{figure}

Figure~\ref{Fig:Correlator2} reveals that $r_{\mathrm{rms}}$ follows two distinct power-law behaviors. For $0 < |U| \le |U_\mathrm{BCS}|$ (with $U_\mathrm{BCS} \sim -t$), we observe a power-law scaling $r_{\mathrm{rms}} \approx r_1 |U/t|^{-\nu_{1}}$ with a small exponent $\nu_1 \sim 0.1$. This behavior is consistent with weakly bound pairs in the BCS regime, where the pair size is large and decreases slowly with increasing $|U|$. As the interaction strength increases, this power-law scaling  gradually breaks down at $|U| = |U_\mathrm{BCS}|$, marking the onset of the BCS–BEC crossover region. For $|U| \ge |U_\mathrm{BEC}|$ (with $U_\mathrm{BEC} \sim -2t$), a second power-law scaling emerges, $r_{\mathrm{rms}} \approx r_2 |U/t|^{-\nu_{2}}$, with an exponent exceeding unity ($\nu_2 > 1$), consistent with the formation of strongly bound pairs. In this BEC regime, the pair size rapidly decreases with increasing $|U|$ once virtual processes of order $\sim t^2/U^2$ become increasingly suppressed, leading to more localized pairs. Thus, for each $n$, the onset of this second power-law behavior at $U_\mathrm{BEC}$ marks the interaction strength at which the system becomes predominantly in the BEC regime: this provides the correlator-based criterion used in Fig.~\ref{Phase_Diagram1} (black squares).

 A similar construction can be perfomed using $|U_{\mathrm{BCS}}|$, thereby identifying the region where BCS pairing dominates. {This also unambiguously delimits the crossover region,}
where {no well-defined power-law scaling is observed.} These results demonstrate that the correlator-based definition of $p(r)$ provides a sensitive and physically transparent probe of pairing across the BCS–BEC crossover in confined systems. Despite its success in identifying the crossover, the correlator-based method does not fully resolve the properties of the insulating phase and the persistence of superfluidity in regions away from the trap center. To address these features, we therefore turn to a more local quantity, namely the single-site entanglement.

\subsection{Persistent superfluid states at edges}

To investigate locally the insulating phase at the trap center and the persistent superfluid edges \cite{sanino2024entanglement}, we analyze the single-site entanglement quantified by the von Neumann entropy 
\begin{align}
    S_j = -\frac{1}{2}\sum_a w_{a,j} \log_2 w_{a,j},
\end{align}
where the probabilities $w_{a,j}$ --- of finding site $j$ in one of the local states: empty ($a=0$), singly occupied with spin-$\uparrow$ or spin-$\downarrow$ ($a=\uparrow, \downarrow$), or the doubly occupied state ($a=2$) --- are obtained from the diagonal elements of the reduced density matrix of site $j$ in the occupation basis. Notice that the prefactor $1/2$ normalizes the entropy such that $S_j^{max}=1$, corresponding to a state in which all four local configurations occur with equal probability.

\begin{figure}[tbh!]
\begin{centering}
\includegraphics[scale=0.38]{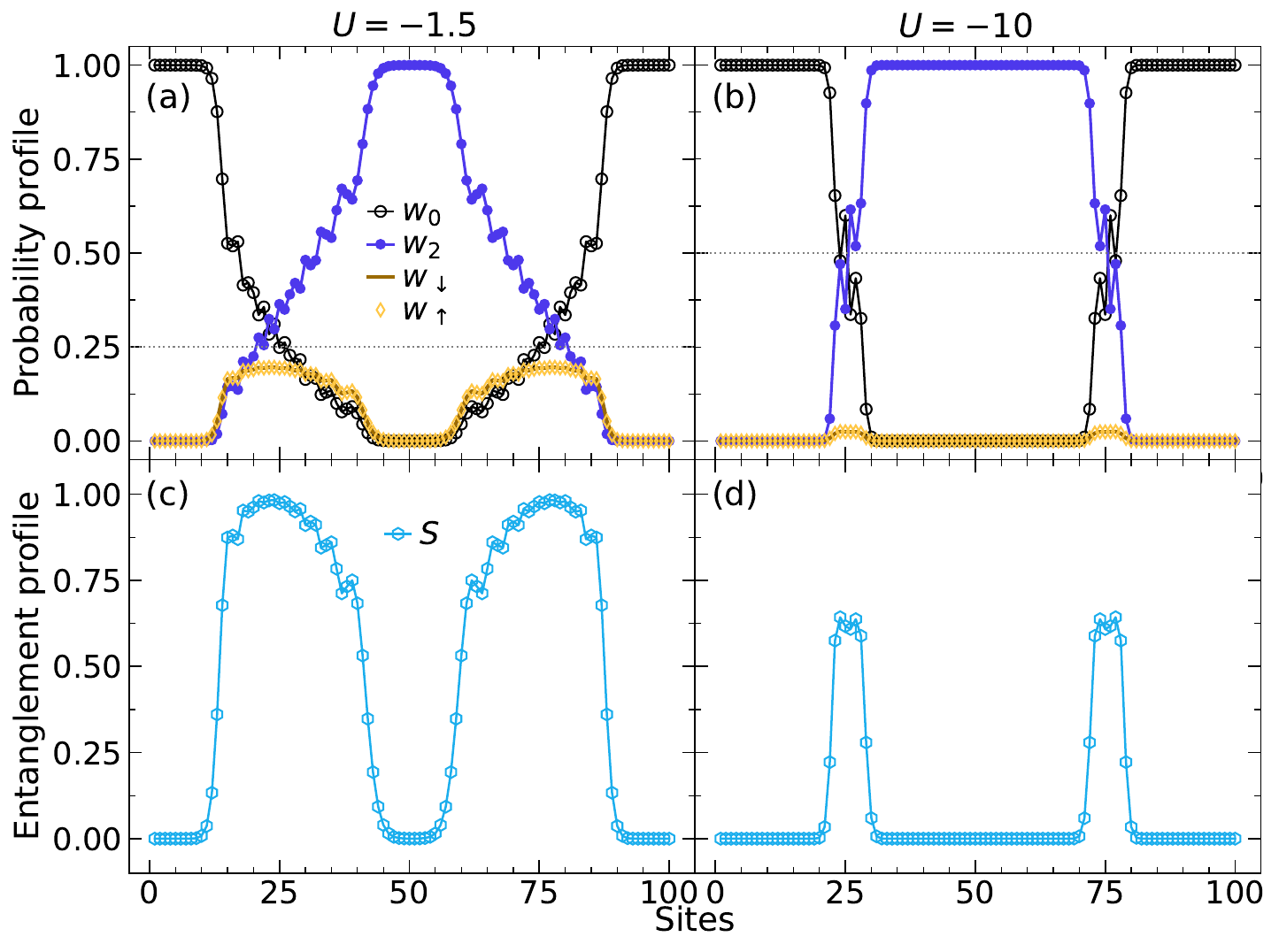}
\par\end{centering}
\caption{Probability profiles (top panels) and entanglement profiles $S_j$ (bottom panels) for the BCS ($U=-1.5t)$ and BEC ($U=-10t$) regimes. In all cases $L=100$, $n=1$, and $k=0.002$.  
\label{fig3}}
\end{figure}

As shown in Figure \ref{fig3}, $S_j$ vanishes when a site is in a pure state of either double occupancy or emptiness, as observed both in the central region and at the outermost sites. In the intermediate region --- the edges --- a persistent superfluid character is observed. In the BEC regime ($U=-10t$), the local states are predominantly composed of empty and doubly occupied configurations, leading to smaller values of $S_j$ ($\approx 0.625$), as previously reported \cite{sanino2024entanglement}. Nevertheless, a small contribution from unpaired electrons, $w_{\uparrow} = w_{\downarrow} \approx 4t^2/|U|^2$, can still be observed, arising from second-order virtual hopping processes, as discussed in the effective models (section II.A). In the BCS regime ($U=-1.5t$), the edge sites display a more balanced distribution among the four local probabilities, thus $S_j$ approaches unity. This is compatible with the mean-field results in Fig.~\ref{MF_WF_k=002}, where the pairing structures are shifted away from the trap center, thereby promoting stronger local mixing at the edges. Therefore, our results demonstrate clear agreement between the DMRG calculations and the predictions of the effective models.

\subsection{Half-chain entanglement as a probe for phase transitions and the BCS-BEC crossover}

\begin{figure}[tbh!]
\begin{centering}
\includegraphics[scale=0.5]{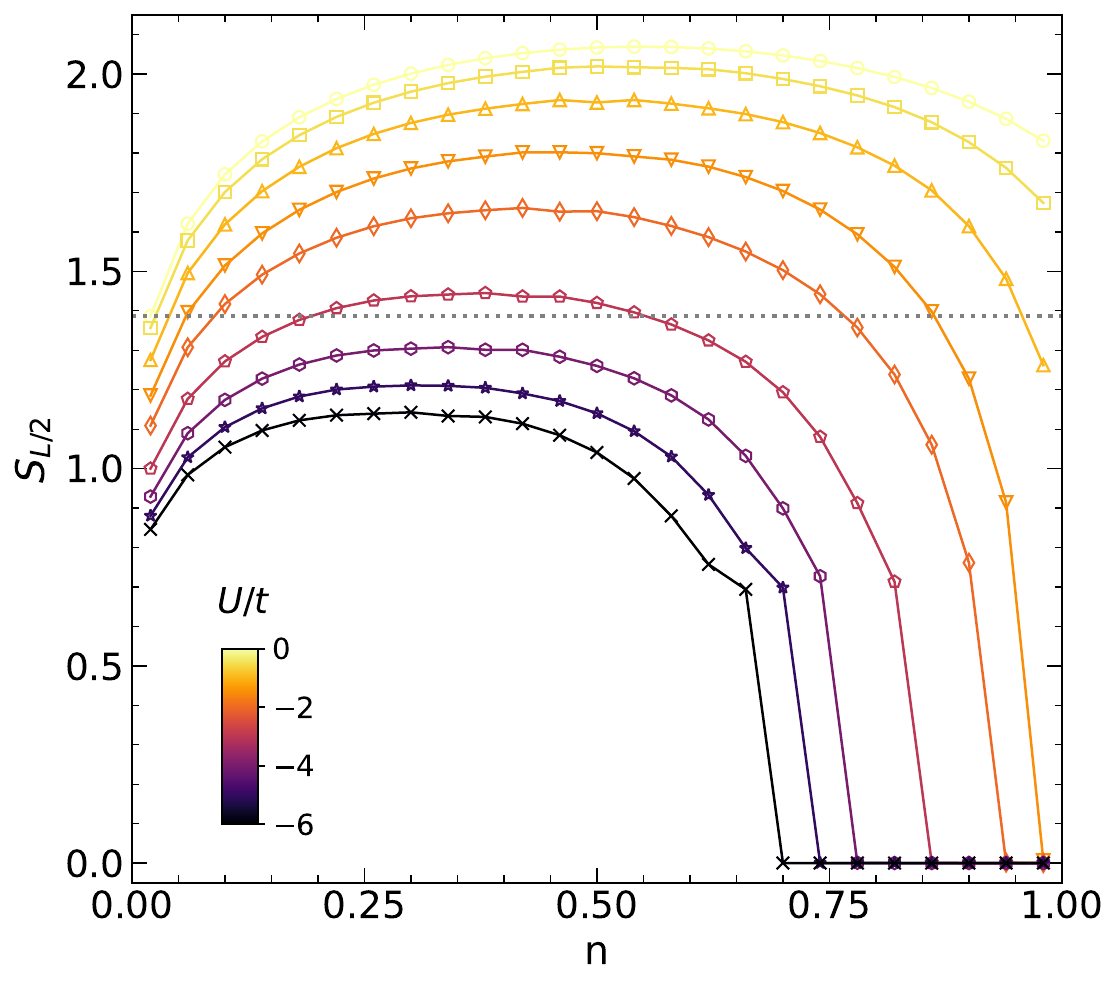}
\par\end{centering}
\caption{$S_{L/2}$ as a function of $n$ for several interaction strengths $U$ (see color legend), with fixed $L = 100$ and $k = 0.002$. The dashed black line indicates the reference value $S_{L/2} = 2 \ln 2$, corresponding to the noninteracting limit, which is used to distinguish between predominantly BEC-like and BCS-like regimes.   
\label{fig_S}}
\end{figure}

We now extend this analysis to a bipartition of the chain into two halves, $L/2$, where the entanglement entropy across the central cut ~\cite{Zawadzki2024_,oliveira2026} directly probes the superfluid–insulator transition,
\begin{align}
    S_{L/2}= -\mathrm{Tr}\!\left(\rho_{L/2} \,\ln \rho_{L/2} \right),
\end{align}
where ${\rho}_{L/2}$ is the half-chain reduced density matrix. Figure~\ref{fig_S} shows $S_{L/2}$ as a function of the density $n$ for different interaction strengths. We observe an abrupt suppression of $S_{L/2}$ precisely at the density for which an insulating region first emerges at the center of the chain, even in cases where this insulating bulk is very narrow. This then signs the emergence of the insulating barrier separating the two halves of the chain and defines the critical density $n_c$ associated with the superfluid--insulator transition. As discussed previously, the increase of the effective confinement $\tilde{k}$ with $|U|$, enables the central region to become fully occupied with fewer particles, thereby decreasing $n_c$ as $|U|$ increases.

For fixed $n$, the overall magnitude of $S_{L/2}$ decreases as $|U|$ increases, providing a more subtle signature of the BCS-BEC crossover than the correlator-based method. Thus, to identify the crossover region, we follow the qualitative behavior suggested by the effective models. At very low densities ($n\approx10^{-2}$), the ground state lies in the BEC regime for any $|U|>0$ \cite{Nozieres1985_}. Consistently, since only tightly bound pairs can cross this boundary, one observes weak entanglement as a consequence of the area law \cite{RevModPhys.82.277}. As $n$ increases, for intermediate values of $|U|$, the system undergoes a crossover from predominantly BEC-like pairing to longer-range BCS-like pairing \cite{Nozieres1985_}. In the BCS regime, $S_{L/2}$ increases as a broader set of configurations contributes across the bipartition, including those involving unpaired electrons in addition to pairs. 

For non-interacting electrons ($U=0$), $S_{L/2}$ is maximal for a given density, since it maximizes the balance among the possible configurations. In the very low-density regime, where the system is effectively occupied by two electrons with opposite spin, $S_{L/2}$ approaches $2\ln 2$ (dashed horizontal line). This value admits an intuitive geometric interpretation that serves as an approximate reference: the system can be seen as two halves connected by a central site. When a single-particle orbital crosses the cut approximately symmetrically, the probabilities of finding the particle in each half become $p_{\rm left}\approx p_{\rm right}\approx 1/2$, yielding an entropy contribution $-p_{\rm left}\ln p_{\rm left}-p_{\rm right}\ln p_{\rm right}\approx\ln 2$~\cite{amico2008}. In the very low-density regime, the confinement localizes the lowest occupied orbitals near the center of the chain, so that the contributions associated with the two spin components ($\uparrow$ and $\downarrow$) add approximately, leading to $S_{L/2}\approx 2\ln 2$.

This geometric picture also provides a simple interpretation of the BCS-BEC crossover. In regimes where particles remain weakly correlated, as in BCS-like behavior, the two spin components contribute approximately independently to $S_{L/2}$, leading to values comparable to or larger than the reference level $S_{L/2}=2\ln 2$. In contrast, in the BEC-like regime, attractive interactions promote the formation of bound pairs that behave effectively as single composite units~\footnote{In 1D, BEC particles behave as spinless fermions, as shown in Ref.~\cite{Girardeau1960_}. Therefore, in the BEC regime the number of degrees of freedom is approximately half that of the $U=0$ case at the same density.}. At most one such pair per bipartition can cross the central link, reducing the number of independent contributions and leading to $S_{L/2}<2\ln 2$. In this sense, the position of the data relative to the dashed reference line in Fig.~\ref{fig_S} provides a simple indication of whether BEC regime becomes dominant.

To substantiate this interpretation, we determine the onset of the BEC regime using both entanglement-based ($S_{L/2} < 2\ln 2$) and the correlator-based methods, finding a very good agreement, as shown in Fig.~\ref{Phase_Diagram1}. Moreover, within this criterion, no BCS–BEC crossover is observed for $|U| > 4t$; in this case, the system undergoes a direct transition from a BEC-like phase to the insulating phase as $n$ increases. Our results not only demonstrate the consistency of the interpretation obtained from the different methods, but also that the half-chain entanglement captures both the superfluid–insulator transition and the BCS–BEC crossover.

\subsection{Phase Diagram}

\begin{figure}[tbh!]
\begin{centering}
\includegraphics[scale= 0.66]{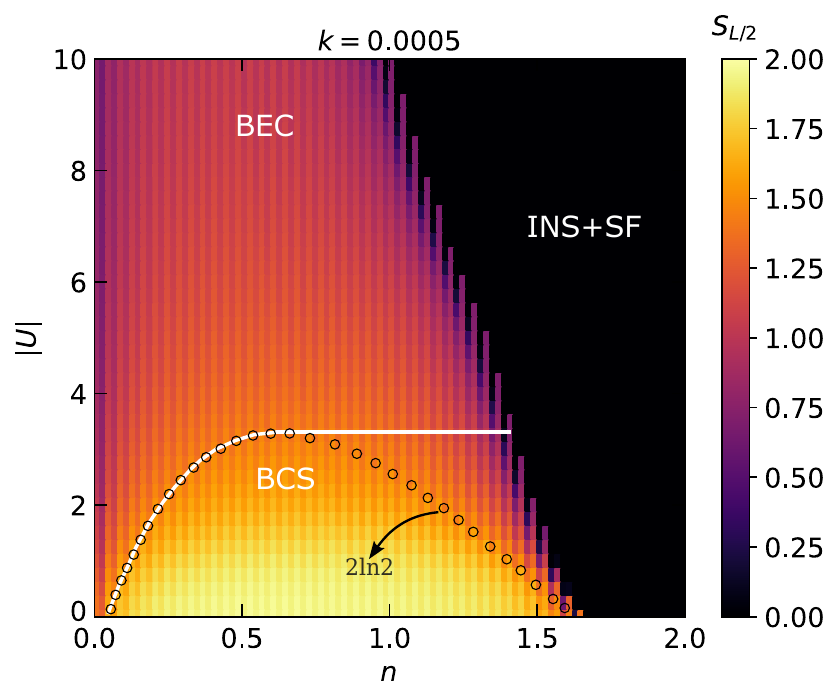}
\includegraphics[scale= 0.66]{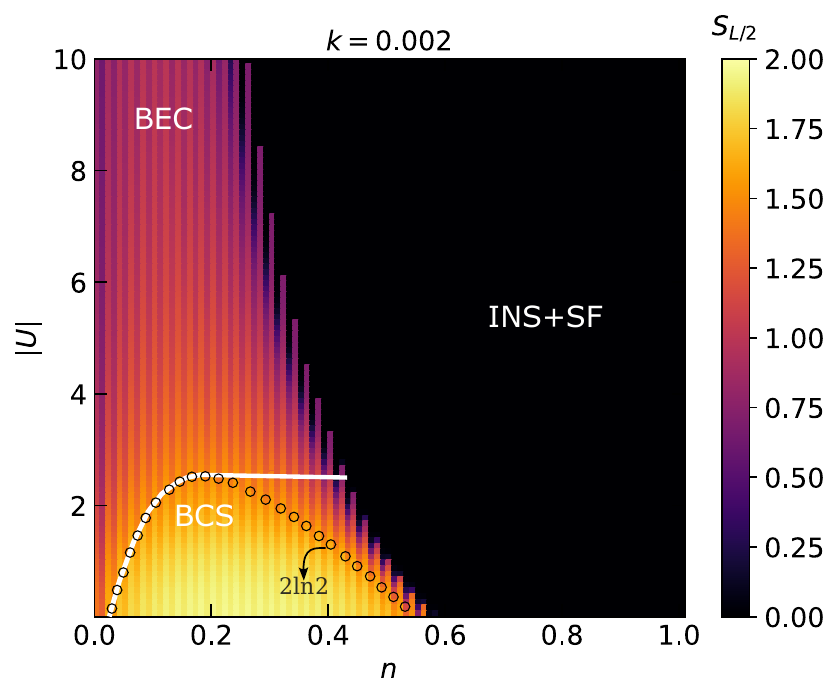}
\par\end{centering}
\caption{Phase diagram as a function of the interaction strength $|U|$ and the average density $n$ for $L=100$ and $k=0.0005$ (top) and $L=200$ and $k=0.002$ (bottom). The diagram is defined by the following criteria: $S_{L/2}<2\ln2$ identifies the low-density BEC-like regime, while $S_{L/2}\ge 2\ln2$ characterizes BCS-like behavior up to a maximum interaction $|U|_{\mathrm{max}}$ at $n = n_{\mathrm{max}}$. For $n>n_{\mathrm{max}}$, the value $|U|=|U|_{\mathrm{max}}$ is kept as the crossover criterion. Finally, $S_{L/2} \le \eta$, with $\eta=0.001$, identifies the phase with a central insulating region surrounded by superfluid (INS+SF).
\label{Phase_Diagram2}}
\end{figure}

To summarize our results, Figure ~\ref{Phase_Diagram2} presents the phase diagram obtained from $S_{L/2}$ for different confinement strengths. The top panel shows close correspondence with the effective predictions of Fig.~\ref{Phase_Diagram1}: the system starts in the BEC regime for $n \ll 1$, crosses over to the BCS regime as $n$ increases (for moderate interaction), and eventually ($n>n_C$) reaches the structure composed by an insulating bulk surrounded by superfluid edges (INS+SF). For $|U| > 4t$ the system undergoes a direct transition from a BEC-like superfluid to the INS+SF phase, in agreement with the effective-model analysis.


{However, Fig.~\ref{fig_S} shows that $S_{L/2}$ increases monotonically with $n$ up to a maximum and then decreases to zero as the system approaches the INS+SF phase. Within the geometric criterion ($S_{L/2}<2\ln2$),  for a given $|U|<4t$ would suggest the onset of a second BEC-like regime at large densities, as indicated by the black circles in Fig.~\ref{Phase_Diagram2} corresponding to $S_{L/2}=2\ln2$. In contrast, analyses based on effective models and correlation functions show no evidence of a second BCS-to-BEC crossover upon increasing $n$; instead, the system evolves from a superfluid BCS regime to an insulating phase with BCS edge states. To resolve this inconsistency, once the condition $S_{L/2}=2\ln2$ reaches a maximum interaction $|U|_\mathrm{max}$, we fix this value for all higher densities. The resulting modified criterion, shown by the white solid line, is in qualitative agreement with the results in Fig.~\ref{Phase_Diagram1}.}



The features discussed above are present for any $k>0$, although the extent of the crossover region depends strongly on $k$, as observed by comparing the top and bottom panels of Fig.~\ref{Phase_Diagram2}. In particular, both the crossover and the superfluid–insulator transition in the bottom panel ($k = 0.002$ and $L = 200$) are shifted toward the left side of the diagram, i.e., toward lower densities and smaller values of $|U|$. This shift has two main origins. First, in strongly confined regimes, increasing $L$ while keeping $k$ fixed has little impact on the ground-state properties, since the filling of the central region is primarily determined by the confinement strength rather than by the total system size. As a result, increasing $L$ effectively shifts the density scale associated with the filling of the central region toward smaller values. Second, increasing $k$ shifts the superfluid–insulator transition toward lower densities. Stronger confinement enhances localization, inducing the transition at smaller fillings and suppressing the formation of long-range BCS pairing (see SM, Sec.~3(c)), thereby shifting the BCS–BEC crossover toward lower values of $|U|$. Consequently, the value of $|U|$ at which the system undergoes a direct transition from the BEC regime to the insulating phase is smaller for $k = 0.002$.

Finally, we recall that the insulating phase consists of a central region dominated by double occupancy, always accompanied by persistent superfluid regions at the edges, giving rise to a composite INS–SF phase. These edge superfluid states exhibit BCS-like character for $|U| < 4t$, while for $|U| > 4t$ they display BEC-like character. The persistent superfluid states remain stable even under strong confinement and at high densities. They are localized around regions where an approximate local particle–hole-like balance develops, as indicated by $w_0 \approx w_2$ in Fig.~\ref{fig3}.

\section{Summary and Conclusions}\label{sec_conclusion}

In summary, our results demonstrate that the effective models provide valuable insight into pairing in the confined 1D Fermi–Hubbard chain. The resulting physical picture, summarized in Fig.~\ref{Phase_Diagram1}, offers a clear interpretation of the DMRG findings. In the strong-coupling regime $|U|\gg t$, the system behaves as hard-core bosonic pairs, thus a simple tight-binding description with renormalized confinement captures the essential physics and reproduces the main features observed in the DMRG results (Fig.~\ref{BEC_Density}).

Moreover, when the central region becomes fully occupied, the system undergoes a transition to a composite insulating-superfluid phase, in which superfluid states persist at the edges. This transition occurs at a critical density $n_c \approx 5.2/\sqrt{\tilde{k}L^2}$ (see Fig.~\ref{n_c}). In strongly confined regimes, the effective repulsion between BEC-like pairs described by Eq.~\eqref{EFF_SCL} expels pairs from the trap center, generating the density oscillations observed in Fig.~\ref{BEC_Density} and reported in Ref.~\cite{sanino2024entanglement}. This mechanism naturally accounts for the persistence of superfluid regions at the edges in the strong-coupling regime$|U|\gg t$.

Confinement also enriches the spatial structure of pairing along the chain. BCS-like pairs can survive at the edges even when the central region become insulating, and are associated with density oscillations observed in Fig.~\ref{MF_WF_k=002}. The same spatial structure is captured in the DMRG results through the probabilities $w_a$ and the single-site entanglement $S_j$ (Fig.~\ref{fig3}). These persistent pairing features at the edges provide a natural explanation for the ``wings'' reported in Ref.~\cite{PhysRevB.82.014202}.

Guided by the effective models, we introduce a method based on a four-particle correlator, defined in Eq.~\eqref{Correlator2}, to characterize the BCS-BEC crossover. To the best of our knowledge, this correlator has not been previously explored in this superconducting context. As shown in Fig.~\ref{Fig:Correlator2}, two distinct power-law scaling regime emerge, corresponding to the BCS and BEC regimes, while the crossover occurs in the intermediate region where no well-defined power law is observed. The second power-law regime emerges for $|U|/t \gtrsim 2$, indicating the continuous shrinking of the pair size with increasing interaction strength, $r_{\mathrm{pair}} \sim |U/t|^{-\nu_2}$. This scaling reflects the approach to the fixed point $|U| \rightarrow \infty$ \cite{Wilson_1974, Wilson1974_RG_} (atomic limit), and its onset signals the emergence of the BEC-like behavior. We further show that the half-chain entanglement identifies not only the superfluid–insulator transition but also provides a useful indicator of the BCS–BEC crossover. These results highlight the usefulness of entanglement-based probes for characterizing many-body phase transitions and crossovers \cite{l8nx-c6nd,Zawadzki2024_,Pauletti2024_}.

Although the identification of the BCS–BEC crossover is not universal and depends on the specific diagnostic employed, the methods introduced here provide a consistent framework to characterize the dominance of each pairing regime within the superconducting phase of confined systems. In particular, the correlation-based approach offers a clear and physically transparent distinction between the BCS and BEC regimes, while also resolving the intermediate crossover region, as illustrated in Fig. \ref{Fig:Correlator2}. The features identified in this work originate primarily from the breaking of translational symmetry, rather than from the specific form of the harmonic confinement. We therefore expect the qualitative behavior reported here to persist in more general trapping potentials.

Finally, our results are qualitatively consistent with the experimental observations of Ref.~\cite{PhysRevLett.111.175302}, where local occupation probabilities were measured at the center of a quasi-1D harmonic trap. In those experiments, an enhanced probability of locally paired states compared to unpaired atoms is associated with BEC-like pairing. As attractive interactions increase, the population of unpaired atoms is suppressed, eventually leading to a transition toward a doubly occupied insulating regime (see Fig.~3 of Ref.~\cite{PhysRevLett.111.175302}). The suppression of unpaired atoms at low densities, followed by their reemergence at higher densisties, is also in qualitative agreement with our results. Our findings further display qualitative similarities with recent 2D experiments, including the increase of the central density with $|U|$ and the emergence of charge-density-wave–like modulations near the trap center with quasi-momenta close to $\pi$ at low temperatures \cite{Hartke2023_,Brown2019_,PhysRevLett.125.113601,PhysRevLett.125.010403,Holten2022_}.  A detailed theoretical analysis of the 2D case, however, is beyond the scope of the present work and is left for future investigation. 


\begin{acknowledgments}

This study was financed in part by the São Paulo Research Foundation (FAPESP), Brazil, under Grants No. 2021/06744-8 and 2023/00510-0. M.~Sanino acknowledges financial support from FAPESP, under Grants No. 2023/02293-7 and 2025/07040-5. G.D. thanks L.N. Oliveira for fruitful discussions. G.D. also thanks the Faculty of Mathematics and Physics at Charles University, Czech Republic, for its kind hospitality.

\end{acknowledgments}

\bibliographystyle{apsrev4-2}
\bibliography{reference}

\newpage

\clearpage
\onecolumngrid

\setcounter{section}{0}
\setcounter{equation}{0}
\setcounter{figure}{0}
\setcounter{table}{0}

\renewcommand{\thesection}{\arabic{section}}
\renewcommand{\theequation}{S\arabic{equation}}
\renewcommand{\thefigure}{S\arabic{figure}}
\renewcommand{\thetable}{S\arabic{table}}

\section*{Supplementary Material}
\addcontentsline{toc}{section}{Supplementary Material}

\section{Atomic limit solution} 

Let us here discuss the case when $t=0$, known as atomic limit. In this limit, the total Hamiltonian can be written as: 

\begin{align}
    H_{0} = \sum_{j} \left[t.k j^2 \left( n_{j,\uparrow} + n_{j,\downarrow} \right) - |U| n_{j,\uparrow} n_{j,\downarrow} \right].
\end{align}

This Hamiltonian is diagonal in the occupation basis and the energy of each site depends only on the number of electrons $n_j$ in each site, which can be written as
\begin{align}
    \epsilon_j(n_j = 0) = 0,
\end{align}
\begin{align}
    \epsilon_j(n_j = 1) = t.kj^2,
\end{align}
\begin{align}
    \epsilon_j(n_j = 2) = 2t.kj^2 - |U|.
\end{align}

In this case, the ground state is obtained by filling the $N$ electrons in the corresponding sectors determined by $j^* = \sqrt{|U|/2kt}$, in accordance with the Pauli principle. As a consequence, the electronic filling in the atomic limit can be divided into three distinct regions along the chain.
{Region I} ($\epsilon_j(2) < \epsilon_j(0) \le \epsilon_j(1)$ with $0 \le j \le j^*$): Favors the occupation of each site by two electrons, decreasing the ground-state energy.
{Region II} ($\epsilon_j(0) \le \epsilon_j(2) < \epsilon_j(1)$ with $j^* < j \le \sqrt{2} j^*$): Still favors two electrons per site, but increases the ground-state energy.
{Region III} ($\epsilon_j(0) < \epsilon_j(1) \le \epsilon_j(2)$ with $\sqrt{2} j^* < j$): Favors only one electron per site, increasing the ground-state energy.

\section{Effective model in strongly coupled limit via Löwdin Perturbation Theory}

For simplicity, let us consider the initial Hamiltonian \eqref{Eq. 1}, but after an translation $j_0 = 0$ in the $|U|\gg t$ regime. Once that is required a high energy to break one electronic pair in this system, we can also extract the effective model via well known Löwdin Perturbation Theory method, that reads: 
\begin{align}
    \tilde H = H_{1} - T_{12} (H_{2} -  E)^{-1} T_{21}.
\end{align}
Here, the sector $1$ referees to low energy paired electrons and the sector $2$ excited states where there are two unpaired electrons. Therefore, we can approximate the contribution from the second sector energy by $H_2 \approx E + |U|$, where $E$ is the ground state energy. 

Using the projector $P_{1}$, which projects the state into the paired sector, and observing that the hoping term is responsible to the transitions between sectors, we can then write: 
\begin{align}
    \tilde H &= \sum_j   \left( -|U| + 2t.kj^2 \right) c_{j,\uparrow}^\dagger c_{j,\downarrow}^\dagger c_{j,\downarrow} c_{j,\uparrow}  -\frac{t^2}{|U|} P_{1} \sum_{j\sigma} (c_{j\sigma}^\dagger c_{j+1\sigma} +\mathrm{h.~c.}) \sum_{j\sigma}  (c_{j\sigma}^\dagger c_{j+1\sigma} +\mathrm{h.~c.}) P_1^\dagger .
\end{align}
The first term only accounts the energy of the paired electrons, where the second  term is an second order contribution of the unpaired sector into the paired sector. 

As the projector $P_1$ guarantees that only terms that don't break electronic pairs survive, it is then straightforward to show that:  
\begin{align}
    \tilde H &= \sum_j \left( -|U| + 2t.kj^2 \right) c_{j,\uparrow}^\dagger c_{j,\downarrow}^\dagger c_{j,\downarrow} c_{j,\uparrow} -\frac{2t^2}{|U|} \sum_{j} (c_{j\uparrow}^\dagger c_{j\downarrow}^\dagger c_{j+1\downarrow} c_{j+1\uparrow} +\mathrm{h.~c.}) -\frac{2t^2}{|U|} N_e + \frac{2t^2}{|U|} \sum_{j\sigma} (c_{j\sigma}^\dagger c_{j\sigma} c_{j+1 \sigma}^\dagger c_{j+1\sigma}).
\end{align}

Now, let us introduce the on-site BEC operator $\phi_j^\dagger = c_{j\uparrow}^\dagger c_{j\downarrow}^\dagger$ to represent the pair at each site $j$. Since the operator $\phi_j^\dagger \phi_j$ denotes the number of BEC quasiparticles at site $j$, each composed of one spin-up and one spin-down particle, in the strongly coupled limit we have $\phi_j^\dagger \phi_j = n_{j\sigma}$. Therefore, we can rewrite the effective model \cite{RevModPhys.62.113} as:
\begin{align}\label{H_AA}
    \tilde H &= \sum_j \left( -|U|\left(1+ \frac{4t^2}{|U|^2} \right) + 2t.kj^2 \right) \phi_{j}^\dagger  \phi_{j} -\frac{2t^2}{|U|} \sum_{j} (\phi_{j}^\dagger \phi_{j+1} +\mathrm{h.~c.})+ \frac{4t^2}{|U|} \sum_{j}\phi_{j}^\dagger \phi_{j} \phi_{j+1}^\dagger \phi_{j+1}.
\end{align}

This effective model describes a hard-core bosonic BEC quasi-particle that can move through virtual processes (singlet states) represented by the hopping term and experiences a nearest-neighbor positive Coulomb interaction. 
The same effective model can also be derived using the Schrieffer–Wolff transformation \cite{RevModPhys.62.113}, and it is known to accurately capture the properties of the system in the strongly coupled limit. 

\section{Effective tight-binding model for BEC quasiparticles}

An interesting case with an easily obtainable analytical solution is the two-site (dimmer) negative-$U$ Fermi-Hubbard model, whose Hamiltonian can be written as:
\begin{align}
     H_{\mathrm{D}} = t.kj^2 (n_1 + n_2) &+ 2V_j n_2 - |U| \left[ n_{1\uparrow} n_{1\downarrow} + n_{2\uparrow} n_{2\downarrow} \right] -t\left[c_{1,\uparrow}^\dagger c_{2,\uparrow}  + c_{1,\downarrow}^\dagger c_{2,\downarrow} + \mathrm{h.~c.} \right] .
\end{align}
Here, $V_j = t.k\left(j+\frac{1}{2}\right)$. Focusing on the two particle singlet sector, the Hamiltonian of the dimmer can be written as 
\begin{eqnarray}
\hat H_{2,0} \equiv 
\begin{pmatrix}
- |U| & -\sqrt{2}t & 0 \\
-\sqrt{2}t  &  0 & -\sqrt{2}t \\
0 & -\sqrt{2}t  &  - |U|
\end{pmatrix} + \begin{pmatrix}
-2V_j & 0 & 0 \\
0  &  0 & 0 \\
0 & 0  &  2V_j
\end{pmatrix}.~~~~
\end{eqnarray}

Now, by defining the states:  
\begin{align}
\ket{2}_{+} =  \frac{1}{\sqrt 2}\left( c_{1,\uparrow}^\dagger c_{1,\downarrow}^\dagger + c_{2,\uparrow}^\dagger c_{2,\downarrow}^\dagger \right)\ket{0};~~
\ket{2}_{-} =  \frac{1}{\sqrt 2}\left( c_{1,\uparrow}^\dagger c_{1,\downarrow}^\dagger - c_{2,\uparrow}^\dagger c_{2,\downarrow}^\dagger \right)\ket{0};~~
\ket{2}_{S} =  \frac{1}{\sqrt 2}\left( c_{1,\uparrow}^\dagger c_{2,\downarrow}^\dagger + c_{2,\uparrow}^\dagger c_{1,\downarrow}^\dagger \right)\ket{0},
\end{align}
and the quantity 
\begin{align}
    &\delta= \frac{2t}{|U|}, 
\end{align}
using perturbation theory (with $V_j \ll t$) the eigenvalues and energies of this sector can be written as: 
\begin{align}
    &\ket{2,0,0} \approx \mathcal{N}^{-1}\left( \ket{2}_{+} + \frac{2\delta}{(1+ \sqrt{1+(2\delta^2)})}\ket{2}_{S}  \right); &\epsilon_{(2,0,0)} \approx 2t.kj^2 + 2V_j -\frac{|U|}{2}\left(\sqrt{1+(2\delta)^2}+1 \right);\\
    &\ket{2,0,1} \approx \ket{2}_{-}; &\epsilon_{(2,0,1)}  \approx \epsilon_{(2,0,0)} + \frac{|U|}{2}\left(\sqrt{1+(2\delta)^2}-1 \right);  \\
    &\ket{2,0,2} \approx  \mathcal{N}^{-1}\left( \ket{2}_{S} - \frac{2\delta}{(1+ \sqrt{1+(2\delta^2)})}\ket{2}_{+}  \right);  &\epsilon_{(2,0,2)}  \approx \epsilon_{(2,0,0)}  +|U| \sqrt{1+(2\delta)^2},
\end{align}
with $\mathcal{N}$ the normalization factor.

Let us first note that for $\delta> 0$, the ground state is the state $\ket{2,0,0}$, which is a symmetric superposition of doubly occupied sites and the vacuum, together with a small contribution of the singlet state. A long chain constructed from such dimers will naturally tend to form a ground state that favors electron pairing in the configuration $\ket{2,0,0}$. In a realistic Fermi–Hubbard chain, however, dimers interact with one another, introducing additional features beyond this simplified picture. Nevertheless, if $|U| \gg t, t.kj^2$, some of the dimmer features are expected to persist. When $\delta \ll 1$, the system behaves similarly to the atomic limit, in which the correlations between neighboring sites tend to zero. Here, however, we consider an intermediate case, where $|U|$ is large enough to ensure that the ground state is composed in majority of paired electrons, while still maintaining an small correlations between neighboring sites along the chain. In this case, paired electrons then act as the quasi-particles of the system.

We can then approximated the lower energy state of the half filed Fermi-Hubbard dimmer as: 
\begin{align}
    \ket{2,0,0} \approx \frac{1}{\sqrt{\left(1+\delta^2\right)}}\left( \ket{2}_S + {\delta} \ket{2}_S \right),
~\mathrm{with~energy}~
    \epsilon_{(2,0,0)} \approx 2t.kj^2 + 2 V_j -|U|(1+\delta^2).
\end{align}

However, in the case of a Hubbard chain, site $j$ can form a dimer either with site $j+1$ or with site $j-1$. Therefore, let us define the contribution of this BEC pair localized at site $j$ as a field operator: 
\begin{align}
    \phi^\dagger_{j} = \frac{1}{\sqrt{\left(1+2\delta^2\right)}} \left( c_{j,\uparrow}^\dagger c_{j,\downarrow}^\dagger + {\delta}  c_{j\uparrow}^\dagger c_{j+1\downarrow}^\dagger +  {\delta}  c_{j-1\uparrow}^\dagger c_{j\downarrow}^\dagger  \right).
\end{align}
The plane wave expansion of this localized pair is
\begin{align}
    \phi^\dagger_{q} = \frac{1}{\sqrt{L}} \sum_j e^{-i\pi q.j/L} \phi_j ^\dagger ,
\end{align}
with $  1 \le q \le L $.

Let us consider this hard bosonic field $\phi_j$ obtained from the dimmer ground state as an quasiparticle in which we can use to find the approximated ground state. In the strongly coupled limit the unperturbed part of the Hamiltonian is the on site contribution, and the hoping is the perturbative term. Therefore, it will be usefully to us to compute:
\begin{align}
     \bra{0} \phi_q \left[-|U| \sum_l n_{l,\uparrow} n_{l,\downarrow} \right] \phi_q^\dagger \ket{0} \approx  -|U|(1-2\delta^2),
\end{align}
and 
\begin{align}
     \bra{0} \phi_q \left[ -t \left(\sum_{l} c_{l,\uparrow}^\dagger c_{l+1,\uparrow} + c_{l,\downarrow}^\dagger c_{l+1,\downarrow} + h.c  \right) \right] \phi_q^\dagger \ket{0}  \approx -2|U|\delta^2({1 + \cos(\pi q/L)}).
\end{align}


The total energy in this simple first order approximation, excluding the harmonic potential, is then: 
\begin{align}\label{Cooper_Energy}
    \varepsilon_q = -|U| - 2|U|\delta^2  \cos(\pi q/L) + \mathcal{O}(\delta^4).
\end{align}

Note that the energy in Eq. \eqref{Cooper_Energy} has the same form as that of a tight-binding model with a effective hooping $\tau = |U|\delta^2$. Therefore, we can write an effective Hamiltonian as
\begin{align}\label{EFF_H}
    H_{BEC} = \sum_j \left( -|U| + 2t.kj^2 \right)\phi_j^\dagger \phi_j - {|U|\delta^2} \sum_j (\phi_j^\dagger \phi_{j+1} + h.c),
\end{align}
or in unities of the hoping: 
\begin{align}\label{EFF_HS}
    H_{BEC} = {|U|\delta^2}\left(\sum_j \tilde k j^2 \phi_j^\dagger \phi_j -\sum_j (\phi_j^\dagger \phi_{j+1} + h.c) \right).
\end{align}
Here, the effective harmonic potential is therefore 
\begin{align}
    \tilde k = \left(\frac{2k.t}{|U|\delta^2} \right).
\end{align}

\subsection{Relation with the effective model obtained via perturbation theory}

Now, we can notice certain similarities between the model deduced here via dimerization and the more accurate model obtained through perturbation theory in section 2. However, despite its simplicity, Eq.~\eqref{H_AA} cannot be easily diagonalized due to the presence of a quartic term. To proceed, we apply Wick’s theorem and obtain a mean-field approximation for this quartic operator as:
\begin{align}
\phi_{j}^\dagger \phi_{j} \phi_{j+1}^\dagger \phi_{j+1}
&\approx \langle \phi_{j}^\dagger \phi_{j} \rangle\, \phi_{j+1}^\dagger \phi_{j+1} + \langle  \phi_{j+1}^\dagger \phi_{j+1} \rangle \phi_{j}^\dagger \phi_{j} \nonumber \\
&
- \langle \phi^\dagger_{j} \phi_{j+1}\rangle\, \phi^\dagger_{j+1} \phi_{j}
- \langle \phi^\dagger_{j+1} \phi_{j}\rangle\, \phi^\dagger_{j} \phi_{j+1} \nonumber \\
&
- \langle  \phi_{j}^\dagger \phi_{j} \rangle \langle  \phi_{j+1}^\dagger \phi_{j+1} \rangle
+ \langle \phi_{j}^\dagger \phi_{j} \rangle \langle \phi_{j+1}^\dagger \phi_{j+1} \rangle.
\end{align}
where non-conservative terms were not considered.

Finally, ignoring the zero energy, the effective mean-field Hamiltonian in strongly coupled limit can be written as: 
\begin{align}
    \tilde H &= \sum_j \left( \frac{4t^2}{|U|} ( \langle \tilde n_{j+1} \rangle + \langle  \tilde n_{j-1}  \rangle  )+ 2t.kj^2 \right) \phi_{j}^\dagger  \phi_{j} -\frac{2t^2}{|U|} \sum_{j} (1+2\langle \phi_{j+1}^\dagger \phi_j \rangle )(\phi_{j}^\dagger \phi_{j+1} +\mathrm{h.~c.}).
\end{align}
Here, $\tilde n = \phi_{j}^\dagger  \phi_{j}$. If $\langle \tilde n_{j+1} \rangle \approx \langle \tilde n_{j-1} \rangle$ and the density is low enough so $ \langle \tilde n_j \rangle < 1$, the pairs can be approximated by dimer solutions. In this situation, $\langle \phi_{j+1}^\dagger \phi_j \rangle \approx 1/2$, and consequently $\tilde H \approx H_{\mathrm{BEC}}$.




\section{Mean-field approach}

Using Wick’s theorem on the interaction term of the Hubbard model, one can express the quartic operator as:
\begin{align}
n_{i\uparrow} n_{i\downarrow}
&\approx \langle n_{i\uparrow}\rangle\, n_{i\downarrow} + n_{i\uparrow}\, \langle n_{i\downarrow}\rangle \nonumber \\
&
- \langle c^\dagger_{i\uparrow} c_{i\downarrow}\rangle\, c^\dagger_{i\downarrow} c_{i\uparrow}
- \langle c^\dagger_{i\downarrow} c_{i\uparrow}\rangle\, c^\dagger_{i\uparrow} c_{i\downarrow} \nonumber \\
&
+ \langle c_{i\downarrow} c_{i\uparrow} \rangle\, c^\dagger_{i\uparrow} c^\dagger_{i\downarrow}
+ \langle c^\dagger_{i\uparrow} c^\dagger_{i\downarrow}\rangle\,  c_{i\downarrow} c_{i\uparrow} \nonumber \\
&
- \langle n_{i\uparrow}\rangle \langle n_{i\downarrow}\rangle
+ \langle c^\dagger_{i\uparrow} c_{i\downarrow}\rangle \langle c^\dagger_{i\downarrow} c_{i\uparrow}\rangle
- \langle c_{i\downarrow}  c_{i\uparrow} \rangle \langle c^\dagger_{i\uparrow} c^\dagger_{i\downarrow}\rangle \,.
\end{align}
Once we consider nonmagnetic solutions, $\langle c^\dagger_{i\uparrow} c_{i\downarrow}\rangle = 0$. 

Now, by substituting the non null terms from the mean-field approximation of the quartic term into the full Hamiltonian, ignoring the terms that renormalizes the ground state energy, we found that: 
\begin{align}
     H_{MF} \equiv \sum_{j \sigma} e_j n_{j\sigma}  - &U \sum_j  \left( \langle c_{j\downarrow}  c_{j\uparrow} \rangle\, c^\dagger_{j\uparrow} c^\dagger_{j\downarrow} +\mathrm{h.~c.} \right)  -t \sum_{j\sigma} \left( c_{j\sigma}^\dagger c_{j+1 \sigma} +\mathrm{h.~c.} \right), 
\end{align}
where 
\begin{align}
    e_j = -\mu + k(j-j_0)^2 - |U| \langle n_j \rangle.
\end{align}

Under the particle-hole transformation $h_{j \downarrow }^\dagger  =  c_{j\downarrow} $  for the electrons in the spin down sector, in order to recovery the usual form of quadratic Hamiltonians, we can rewrite the total Hamiltonian in the usual form as:
\begin{align}
     H_{MF} \equiv \mathcal{E}_{0}  + \sum_{j} e_j c_{j\uparrow}^\dagger c_{j\uparrow}-t\sum_{j} \left( c_{j\uparrow}^\dagger c_{j+1 \uparrow} +\mathrm{h.~c.} \right)  &+ \sum_{j} (-e_j) h_{j\downarrow}^\dagger h_{j\downarrow}  +t \sum_{j} \left( h_{j\downarrow}^\dagger h_{j+1\downarrow} +\mathrm{h.~c.} \right)  \nonumber \\ 
     +&|U| \sum_j  \left( \langle c_{j\downarrow} c_{j\uparrow} \rangle\, c^\dagger_{j\uparrow} h_{j\downarrow} +\mathrm{h.~c.} \right),
\end{align}
where 
the "vacuum" energy of this mean-filed model is: 
\begin{align}
     \mathcal{E}_{0} = ~&|U| \sum_j \left(\langle n_{j}\rangle^2 - \langle c^\dagger_{j\uparrow} c^\dagger_{j\downarrow}\rangle^2 \right) + \sum_j e_j.
\end{align}
This mean-field approximated Hamiltonian requires an iterative diagonalization and update of the parameters, up to convergence, to found its ground state properties. Despite this model is hard to be diagonalized analytically, it can be easily done numerically as it is quadratic.

The numerical procedure is performed as follows. First, we consider $\langle c_{j\downarrow} c_{j\uparrow} \rangle = 0$ and a uniform density distribution, and we perform the variational method numerically to update only the density until convergence is reached, with a relative error at each site smaller than $10^{-6}$. Next, we use this optimized density profile and the corresponding chemical potential as a starting point for the non-trivial BCS solutions. We choose an initial seed value $\langle c_{j\downarrow} c_{j\uparrow} \rangle = 0.01$ and perform the variational method by updating both the chemical potential, to keep the deviation in the number of electrons below the threshold of $10^{-6}$, and $\langle c_{j\downarrow} c_{j\uparrow} \rangle$ until it converges with an error below $10^{-6}$ at all sites of the chain.




After diagonalization and iterative procedure up to convergence, it can be written as 
\begin{align}
    H_{MF} = \mathcal{E}_0 + \sum_l \left[ \xi_l \gamma^\dagger_{l,+} \gamma_{l,+} - \xi_l \gamma^\dagger_{l,-} \gamma_{l,-} \right]. 
\end{align}
Here, the fermionic operator $\gamma^\dagger_{l,\pm}$ creates a Bogoliubov quasiparticle (BQP) with energy $\pm \xi_l$, where $\xi_l \ge 0$, which is a combination of an electron with spin $\uparrow$ and a hole with spin $\downarrow$, given by
\begin{align}
    \gamma_{l,-}^\dagger  = \sum_j \left[ \alpha_{l,j} c_{j,\uparrow}^\dagger + \beta_{l,j} h_{j,\downarrow}^\dagger \right],
\end{align}
and 
\begin{align}
    \gamma_{l,+}^\dagger  = \sum_j \left[ -\beta_{l,j} c_{j,\uparrow}^\dagger + \alpha_{l,j} h_{j,\downarrow}^\dagger \right].
\end{align}
These coefficients satisfy $ \sum_j |\alpha_{l,j}|^2 + |\beta_{l,j}|^2 = 1$.



\subsection{Enhance of the confinement in the weakly interaction limit}

Now, note that when the confinement is present, the density profile will be high near the center and drops to zero away from the center of the trap. In this situation, imposing even symmetry of the density distribution, it is reasonable to expect that the density profile around central sites could be represented by Taylor series in second order as following: 
\begin{align*}
    \langle n_j\rangle \approx n_{j_0} + \frac{1}{2} \left(\frac{d^2 n_j}{dj^2}\right)_{j_0}(j-j_0)^2.
\end{align*} 
Therefore, using the above approximation to expand the Hartree--Fock contribution 
$U\langle n_{j\bar{\sigma}} \rangle n_{j\sigma}$ for weakly interaction regimes ($|U| \ll t$), it is straightforward to see that one obtains an effective confinement of the form
\begin{align}
    \tilde k = k + \frac{U}{2} \left(\frac{d^2 n_j}{dj^2}\right)_{j_0}.
\end{align}
Once both $U$ and the second derivative of the density profile are negative, $\tilde{k} \ge k$, explaining the increase of the confinement even for small values of $|U|$.

\subsection{Complementary mean-field results}

\subsubsection{Influence of the interaction strength and Confinement effect}

\begin{figure}[tbh!]
\begin{centering}
\includegraphics[scale=0.55]{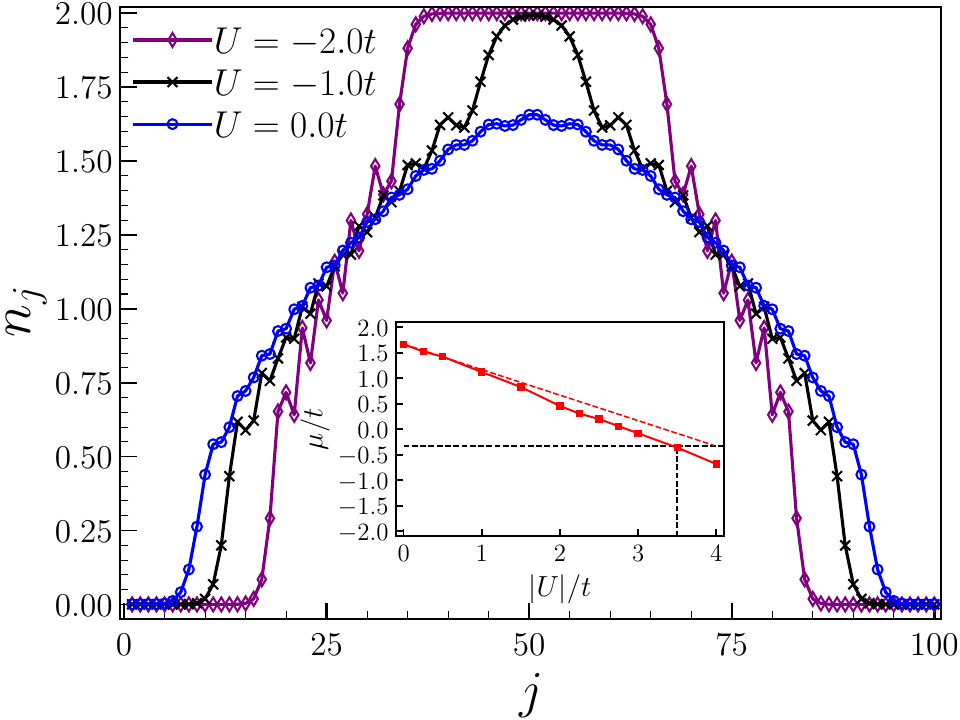}
\includegraphics[scale=0.55]{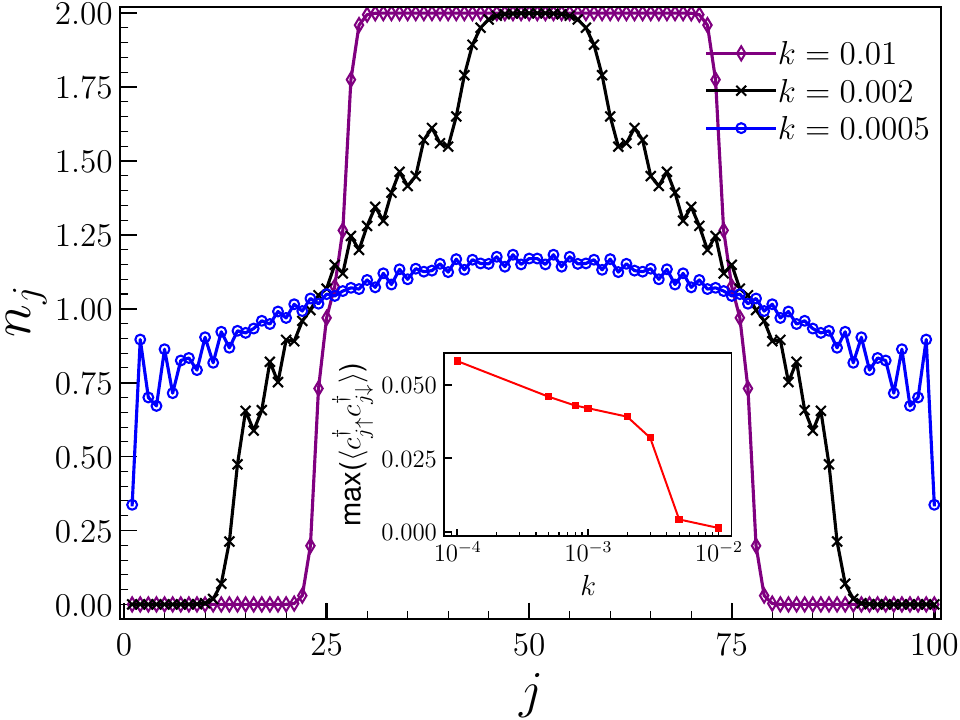}
\par\end{centering}
\caption{(LEFT) Particle distribution $n_j = \langle \sum_\sigma c_{j\sigma}^\dagger c_{j\sigma}\rangle$ along the chain for different values of $U$ (blue curve: $U = 0.0t$, black curve: $U = -1.0t $, and purple curve: $U = -2.0t$). There results where computed with fixed $k=0.002$ and $n=1$. The inset plot shows the behavior of the chemical potential $\mu$ as $|U|$ increases. The dashed red line indicates the expected decrease of $\mu$ with $-|U|/2$ in the homogeneous case. However, when $|U| > 2t$, a faster drop of the chemical potential is observed, falling below $2t$ from its initial value at $U \approx -3.5t$, as indicated by the black dashed lines. (RIGHT) Particle distribution along the chain for $k = 0.0005$ (blue dots),$k = 0.002$ (black dots) and $k = 0.01$) (red dots). There results where computed with fixed $U=-1.25t$ and $n=1$.  The inset plot shows the maximum value of order parameter $\langle  c_{j\uparrow}^\dagger c_{j\downarrow}^\dagger\rangle$ along the chain as function of the confinement. 
\label{MF_k=0.002_Density}}
\end{figure}

As shown in Fig.~\ref{MF_k=0.002_Density} (LEFT), as $|U|$ increases the particle density along the chain becomes increasingly concentrated near the center, eventually forming a smaller, effectively fully filled region. This behavior is consistent with the enhancement of the effective confinement strength for $|U|>0$. To understand this, let us first note that from the mean-field Hamiltonian (see Eq.~(3)), when confinement is present, the density near the center of the chain can be expanded in a Taylor series as shown in S4. This term effectively renormalizes the confinement strength by $\tilde k = k + \frac{U}{2} \left(\frac{d^2 n_j}{dj^2}\right)_{j_0}$. Once both $U$ and the second derivative of the density profile are negative, $\tilde{k} \ge k$, explaining the increase of the confinement even for small values of $|U|$.


In addition, the oscillations between the empty sites and the central region become more pronounced as $|U|$ increases. Figure ~\ref{MF_Spectrum} shows that in strongly confined chains the formation of BQPs extends into the vicinity of the localized states, indicating the formation of pairs far from the center of the chain. This is particularly noteworthy because it indicates that pairs can either spread out from the center (near-gap BCS particles) or spreads way from it, at the interface between regions of higher and lower particle concentration. This can be further inferred from the BQP decompositions shown in Fig.~\ref{MF_WF_k=002}, and these edge-paired electrons are responsible for the strong oscillations observed in the density profile shown in Fig. \ref{MF_k=0.002_Density}.

The pair formation consequence of increasing $|U|$ causes a drop in the chemical potential, as shown in the inset plot of Fig.~\ref{MF_k=0.002_Density}. From the homogeneous result, the decrease in the chemical potential is expected to follow $\mu = \mu_0 - n|U|/2$ (with $n = 1$, represented by the red dashed line). However, unlike the homogeneous case, as $|U|$ increases the density profile along the chain also changes, as shown in the main panel of Fig.~\ref{MF_k=0.002_Density}. Therefore, as $|U|$ increases and the central region becomes more filled, the chemical potential drops faster than its homogeneous counterpart, showing signs of BEC pairing at smaller values ($U \approx -3.5t$) than those expected in the traditional homogeneous BCS-BEC crossover.

As consequence of the confinement, particles are more accumulated near the center of the chain. Figure~\ref{MF_k=0.002_Density} (RIGHT) illustrates this effect through the particle distribution for different values of $k$. For large values of $k$ (purple curve), the confinement becomes too strong, and particles become increasingly localized near the center of the chain, effectively creating a shorter chain with higher density. This localization breaks the momentum-space correlations, destroying the ideal conditions for Cooper pair formation, closing the superconducting gap $\sim \left|U\mathrm{max}(\langle c_{j\uparrow}^\dagger c_{j\downarrow}^\dagger\rangle)\right|$ as shown in inset plot of Fig.~\ref{MF_k=0.002_Density} (RIGHT).


\subsubsection{BCS-BEC crossover from Eq. (12)}

\begin{figure}[tbh!]
\begin{centering}
\includegraphics[scale=0.575]{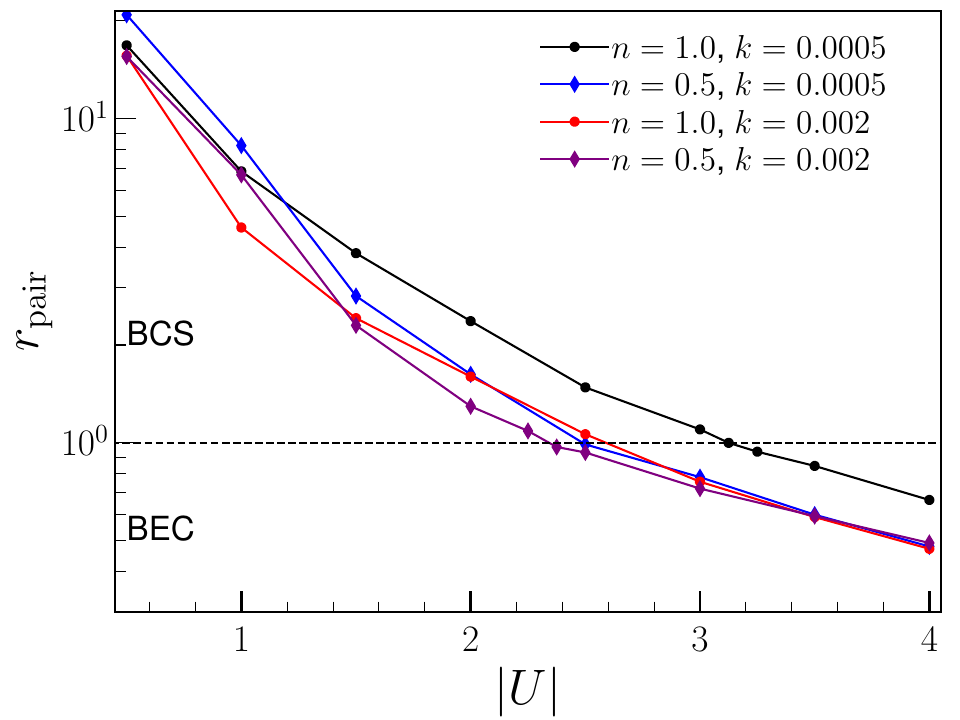}
\par\end{centering}
\caption{ Pair size $r_{\mathrm{pair}}$ as a function of $|U|$ for different fillings and confinement strengths (see legend). As $|U|$ increases, $r_{\mathrm{pair}}$ decreases from the BCS regime ($r_{\mathrm{pair}} \gg 1$) to the BEC regime ($r_{\mathrm{pair}} \le 1$). The horizontal line marks the $r_{\mathrm{pair}}=1$.
\label{r_pairs}}
\end{figure}

The interpretation of Eq.~\eqref{pair_size_estimative} is straightforward: it provides the average distance between spin-up and spin-down particles along the chain. If $r_{\mathrm{pair}} \gg 1$, the system lies in the BCS regime, whereas if $r_{\mathrm{pair}} \le 1$, the system is in the BEC regime. This behavior can be clearly observed in Fig.~\ref{r_pairs}, where the pair size is shown for different fillings, confinement strengths, and interaction values (values indicates in the legend). The results are unambiguous: as $|U|$ increases, the pair size shrinks from values much larger than the intersite spacing (BCS regime) to a tightly bound on-site pair (BEC regime). The horizontal line indicates the criterion $r_{\mathrm{pair}} = 1$ used to separate these two regimes.

\section{Tight-binding under harmonic potential}

Let us start with the initial spinless Hamiltonian, that describes an confined tight-binding, written as 
\begin{align}\label{D_0}
    H = t \sum_j \left[ -(c_j^\dagger c_{j+1} + h.c) + k j^2 c_j^\dagger c_j \right].
\end{align}
This Hamiltonian is diagonalized if is possible to find a linear combination of the operators such as
\begin{align}\label{D_1}
    g_l^\dagger = \sum_j u_{l,j} c_j^\dagger,
\end{align}
which transform the Hamiltonian into 
\begin{align}\label{D_2}
    H = \sum_l \epsilon_l g_l^\dagger g_l.
\end{align}

It is straightforward to show that using the Eqs. \eqref{D_0}, \eqref{D_1} and \eqref{D_2}, far from the edges of the chain, we found out an coupled equation for the energies and coefficients as 
\begin{align}\label{eigenvalue/vector_equation}
    \epsilon_l ~ u_{l,j} =  k j^2 ~ u_{l,j} -( u_{l,j+1} + u_{l,j-1}),
\end{align}
or manipulating the above equation a little bit we find
\begin{align}\label{Aux_1}
    [\epsilon_l +  2] u_{l,j}  =   kj^2 ~ u_{l,j}  -\left[  u_{l,j+1} -2 u_{l,j} + u_{l,j-1} \right]. 
\end{align}

Finite difference methods teach us that the second term on the right-hand side of Eq. \eqref{Aux_1} is the discretized version of the second derivative. By replacing $j$ with $x$ and $u_{l,j}$ with $\Psi_l(x)$, we can then write
\begin{align}
    \epsilon_l \psi_l(x) = \left[ k x^2 - \partial_x^2 \right] \psi_l(x),
\end{align}
which is our friendly and familiar look quantum harmonic oscillator for $\hbar = 1$, $m=1/2$ and $\omega^2 = 4 k$. The quantum harmonic oscillation solution is already very well known as:
\begin{align}
    \psi_l(x) = \frac{1}{\sqrt{(2^l l!)}} \left(\frac{m\omega}{\pi \hbar}\right)^{1/4} \mathcal{H}_l\left( \sqrt\frac{m\omega}{\hbar} x\right) \exp(-\frac{m\omega x^2}{2\hbar}),
\end{align}
where $\mathcal{H}_l(x)$ is the $l$-th Hermite polynomial.

\begin{figure}[tbh!]
\begin{centering}
\includegraphics[scale=0.47]{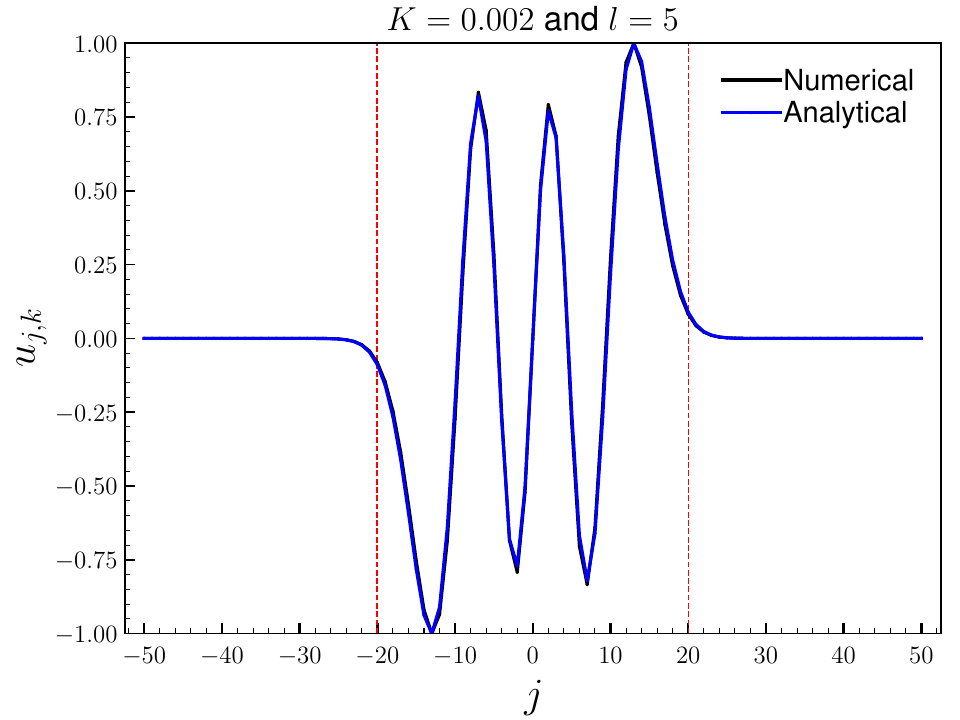}
\includegraphics[scale=0.47]{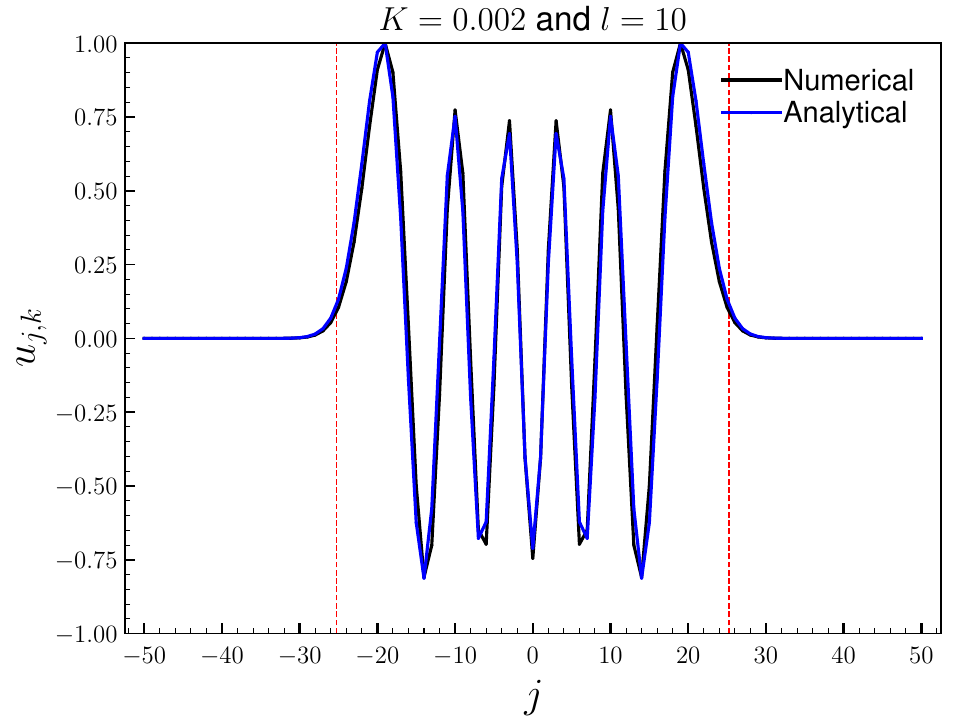}
\end{centering}
\caption{Numerical coefficients normalized by its maximum value computed for $k=0.002$, $L = 101$, and different eigenstates $l$ (values are given in the tittle of each plot). 
\label{Localized_states_1}}
\end{figure}

Returning to the discretized chain notation we can express the eigenvector coefficient as
\begin{align}\label{Analy_Coeff}
    u_{l,j}^{(0)} \approx \frac{A_l}{\sqrt{(2^l l!)}} \left(\frac{\sqrt{k}}{\pi}\right)^{1/4} \mathcal{H}_l\left( \tilde x_j\right) \exp(-\frac{\tilde x_j^2}{2} ).
\end{align}
Here, $\tilde x_j = {k}^{0.25}  j$. The amplitude $A_l$ is in general non unitary here once the chain is limited from $-L/2$ to $+L/2$. Fig. \ref{Localized_states_1} shows the analytical and numerical agreement of the low energy states. In addition, the energies can be expressed as: 
\begin{align}
    \epsilon_l^{(0)} \approx -2 + 2\sqrt{k} (l + 1/2).
\end{align}

\begin{figure}[tbh!]
\begin{centering}
\includegraphics[scale=0.47]{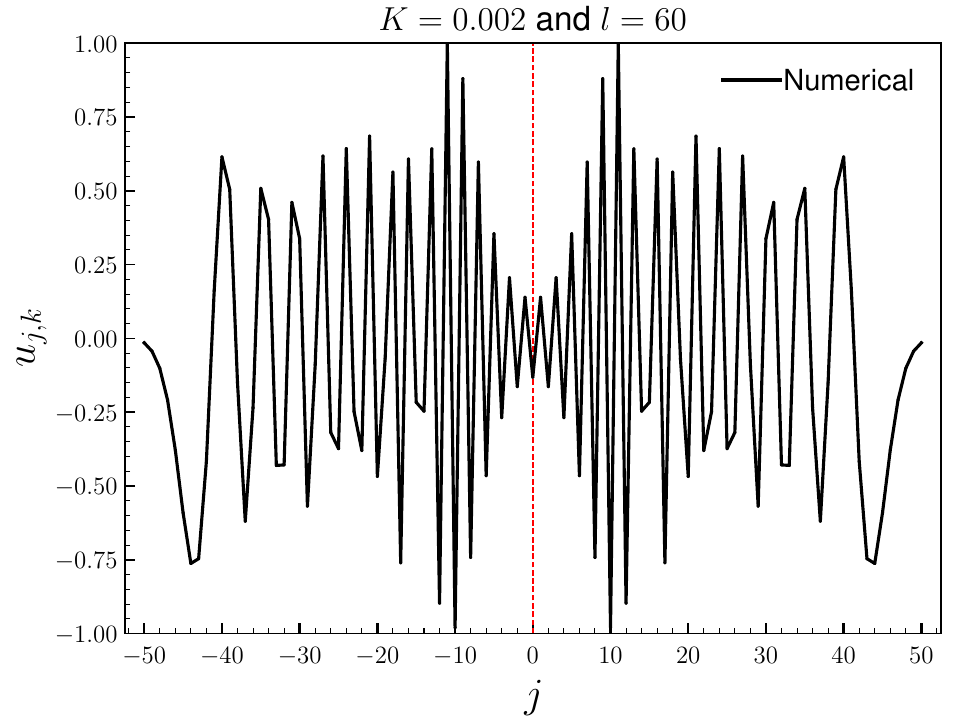}
\includegraphics[scale=0.47]{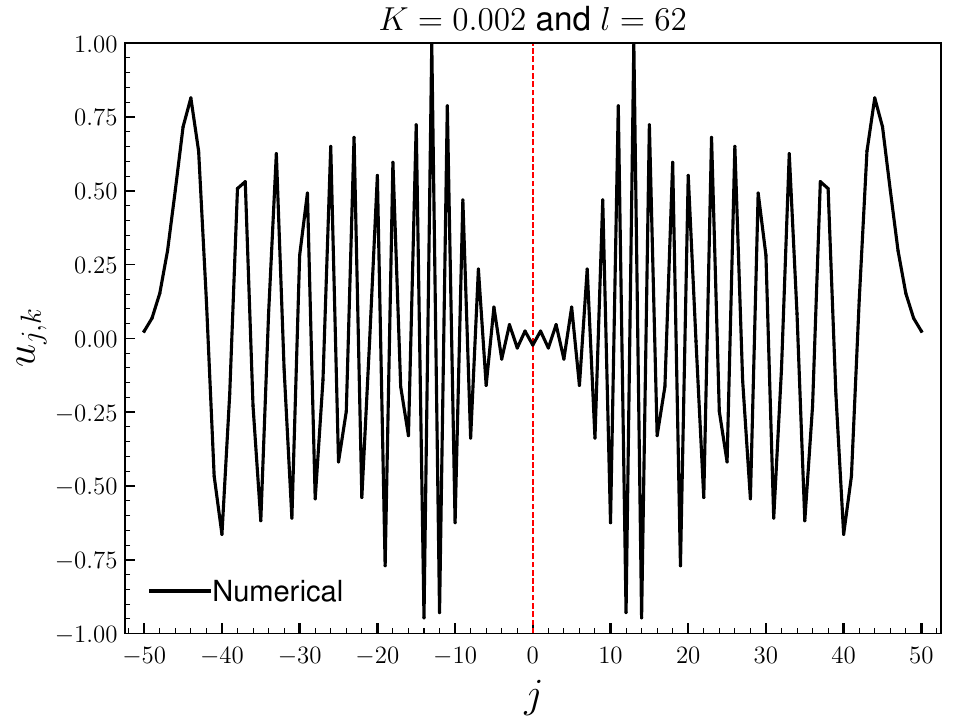}
\end{centering}
\caption{Numerical coefficients normalized by its absolute maximum value computed for $k=0.002$, $L = 101$, and different eigenstates $l$ (values are given in the tittle of each plot). Note the localization near $j^* \approx 30$, with minimum overlap with the central sites, in agreement with the expression Eq. \eqref{D_10} ($j^* \approx 29$).
\label{Localized_states_2}}
\end{figure}

The ground state of $N$ particles in these harmonically confined chains is therefore composed of filling the $N$ lowest energy states. However, as more energetic harmonic states begin to be occupied, increasingly fast oscillations appear in their wavefunctions. When these oscillations acquire a wavelength comparable to the intersite spacing of the chain, the finite-difference approximation used in Eq.~\eqref{Aux_1} breaks down, the corresponding states perceive a discrete translational symmetry instead of a continuous one, and the wavefunctions begin to localize. Figure~\eqref{Localized_states_2} summarizes this discussion with the numerical results.

This breakdown is predicted to occurs around $\epsilon_{l^*} \sim 2$, corresponding to $l^* \sim 2.0/\sqrt{k}$. Numerically we find this value to be $l^* \approx 2.6/\sqrt{k} $. Usually, the number $l$ alone does not contain information about spatial distribution on the chain; nevertheless, due to the localization, this indicates that the wavefunctions start to become spread out around the position $j^* \approx l^*/2$ (because the even symmetry of the confinement potential), instead of being concentrated at the center as in the harmonic states. Therefore, we can estimate $j^*$ as
\begin{align}\label{D_10}
    j^*_{\pm} = \pm \frac{1.3}{\sqrt{k}}.
\end{align}

\begin{figure}[!ht]
  \centering 
  \includegraphics[scale=0.55]{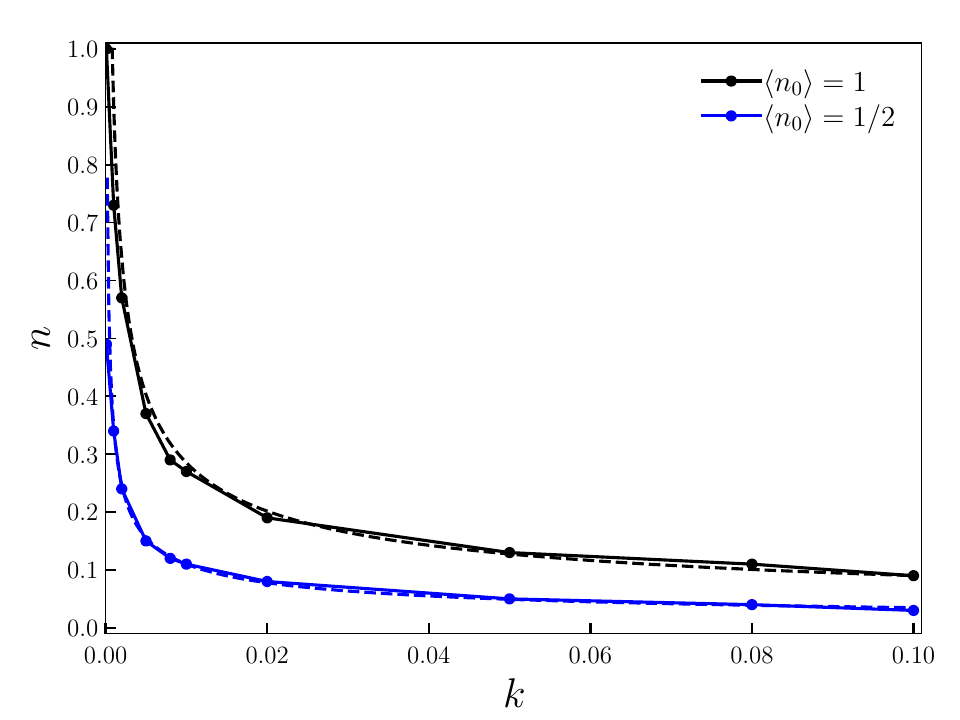}
  \caption{"Critical" densities at the central site as function of $k$ computed numerically (black for $n_0 = 1$ and blue dots for $n_0=1/2$) and with our expression (dashed lines) using $L=101$.}\label{critical_density}
\end{figure}

As $l$ increases above $l^*$, the eigenstates become progressively more localized around the site $j_\pm = \pm l/2$, contributing even less to the density at the center of the chain. Therefore, if the density is high enough that these localized states around $j^*$ begin to be filled, the density at the center of the chain will necessarily approach near unity, indicating a transition to insulating behavior. This concept is illustrated in Fig.~\ref{critical_density}, where the numerical conditions for the central site to approach unity and half occupation are shown, alongside the predicted results based on the computation of $j^*$.

\begin{figure}[tbh!]
\begin{centering}
\includegraphics[scale=0.9]{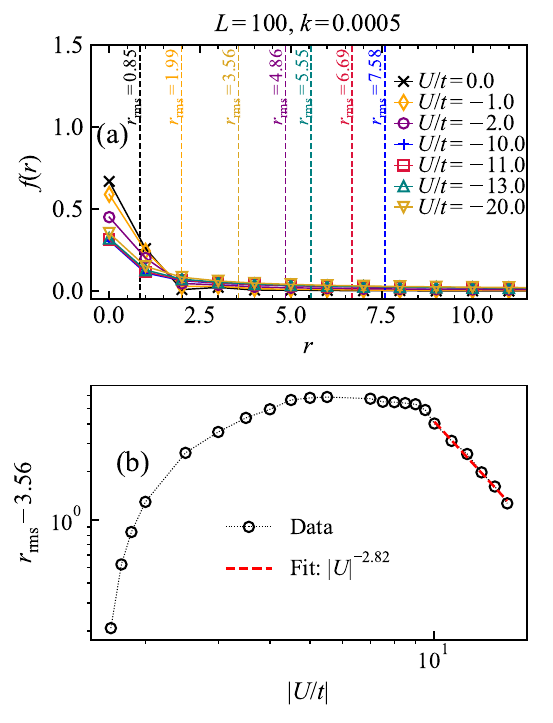}
\par\end{centering}
\caption{Panel (a): Probability distribution $f(r)$ as function of $r$ for different values of $U$. The results were obtained via DMRG at half-filling with $L=100$ and $k = 0.0005$. The values of $U$ are indicated by the legend. (b) R.m.s distance of $f(r)$, $r_{\mathrm{rms}} = \sqrt{\sum_r r^2 f(r)}$, as a function of $U$.
\label{Correlation_Func1}}
\end{figure}

\section{Complementary DMRG result}

\subsection{Correlation functions}

An common approach in the homogeneous scenarios involves computing the pair-pair correlation function
\begin{align}\label{Corre. fun. 1}
    \mathcal{F}_{j}(r) = \langle c_{j+r\uparrow}^\dagger c_{j+r\downarrow}^\dagger c_{j\downarrow} c_{j\uparrow} \rangle, 
\end{align}
which essentially measures the probability that a pair of electrons at site $j$ is correlated with an empty state at site $j+r$, with $r \ge 0$. In simpler terms, this gives us an indication of how far a paired electrons spreads out along the chain.

For instance, it is usefully to define $ f(r)  = \frac{\sum_{j} \mathcal{F}_{j}(r)}{\sum_{j,r'} \mathcal{F}_{j}(r')}$. This quantity results in a probability density for a pair of electrons at a certain site is correlated with an empty state at a distance $r$ along the chain. In homogeneous scenarios, $f(r)$ provides hints about the BCS-BEC regime: an oscillatory behavior with $r$ appears in the BCS regime, whereas a monotonic decay is observed in BEC. However, in the presence of a confinement, we do not observe a clear distinction as $U$ changes (see Fig.~\ref{Correlation_Func1}). Nevertheless, it still provides plenty of useful information.

The panel (a) of Fig.~\ref{Correlation_Func1} shows $f(r)$ as a function of the separation $r$. The plot (b) shows the r. m. s. (root mean square) distance $r_\mathrm{rms}$ of $f(r)$ as function of $U$. We observe that for small values of $|U|$, a sharp peak around $r = 0$ appears in the probability distribution. This reflects the absence of coherent pair formation. In this regime, the only contribution comes from the trivial Hartree–Fock. As $|U|$ increases, coherent pairs begin to form and extend along the chain—first in the BCS regime and later in the BEC regime. The coherence length of the pair increases with $|U|$, note the peak in $r_{\mathrm{rms}}$ around $U = -8t$ in Plot (b). However, once in the BEC regime ($|U| \gg t$), increasing $|U|$ further weakens the effective hopping ($\sim 4t^2/|U|$). This effectively increases the confinement strength ($\tilde{k} \approx k|U|/2t$) and gradually suppresses long-range coherence, causing the average pair mobility to shrink, as observed in the decay of $r_\mathrm{rms}$ shown in Fig. \ref{Correlation_Func1}. An additional interesting observation is that this decay appears to follow a power-law behavior as the system approaches the atomic limit ($U \rightarrow -\infty$).

Another common approach involves computing the pair correlation function for opposite-spins \cite{PhysRevA.91.043612,Strinati2018_BCSBEC_review}
\begin{align}
    \mathcal{G}_{j}(r) = \langle n_{j+r\uparrow} n_{j\downarrow} \rangle - \langle n_{j+r\uparrow} \rangle \langle n_{ j\downarrow} \rangle,
\end{align}
which gives the probability that an electron down at site $j$ and an electron up at site $j+r$ are correlated. Nevertheless, our results show no significant distinction between the BEC/BCS phases in confinement regimes. 

\subsection{Entanglement and charge gap}

\begin{figure}[tbh!]
\begin{centering}
\includegraphics[scale=0.48]{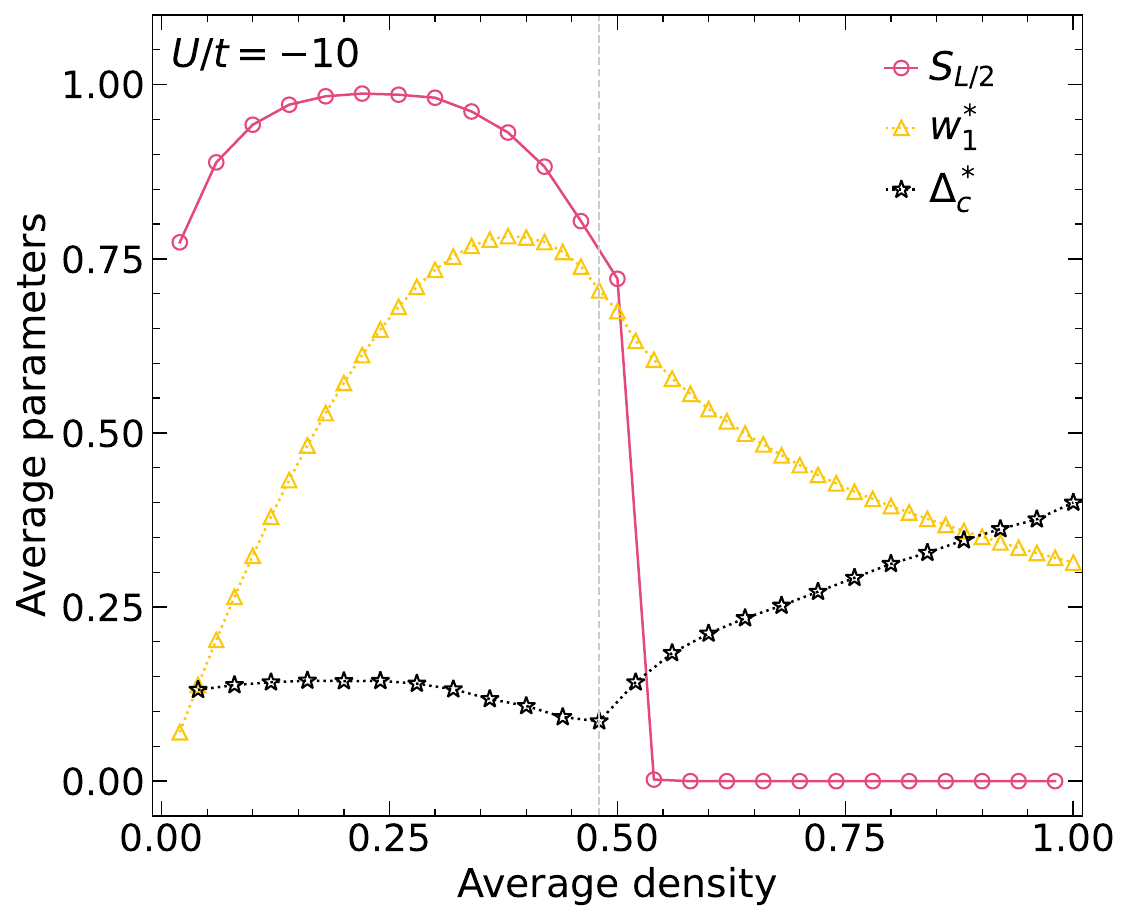}
\includegraphics[scale=0.45]{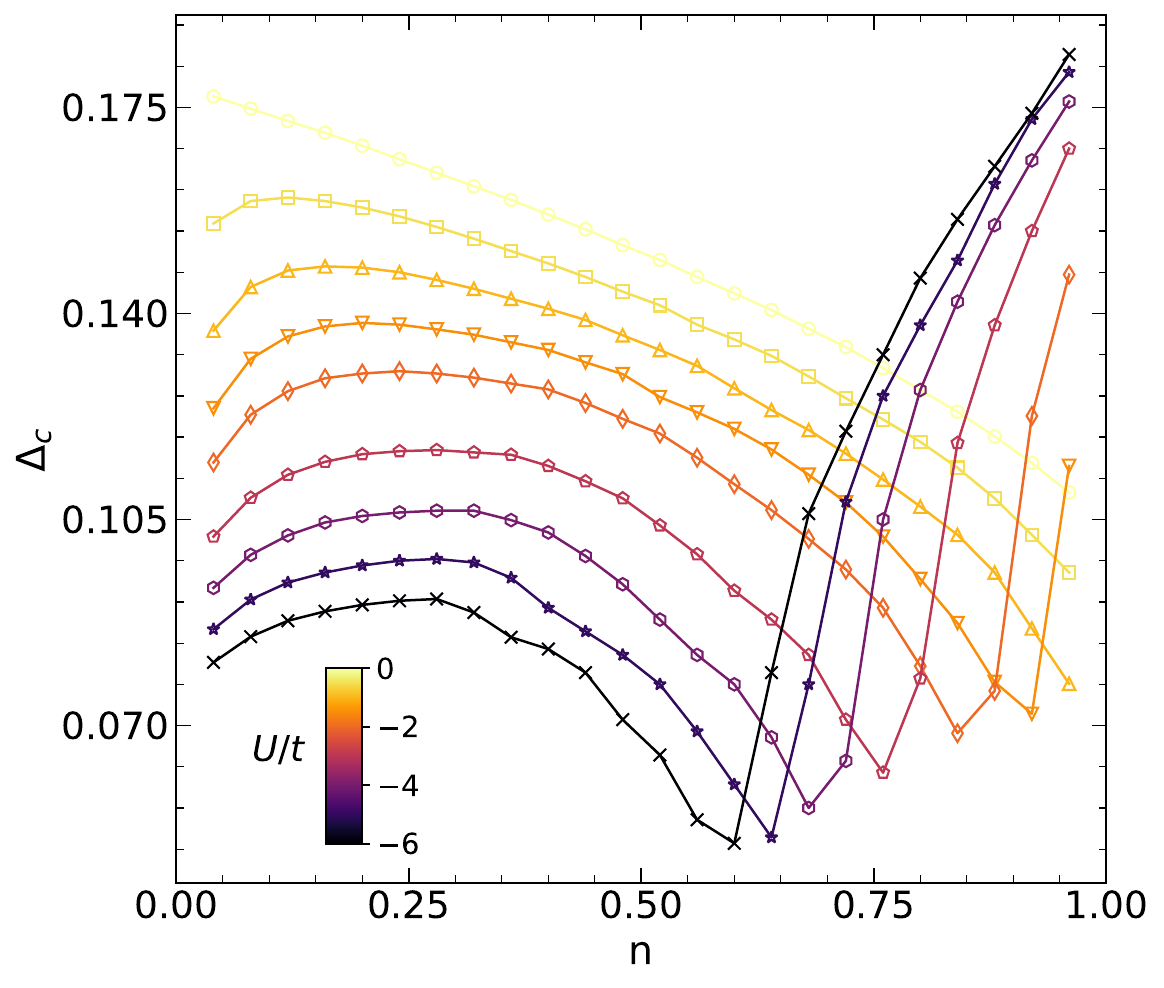}
\par\end{centering}
\caption{(LEFT) Average quantities detecting phase transitions. In the BEC regime, half-chain entanglement $S_{L/2}$, average single occupancy $w_1$ and charge gap $\Delta_c$ are shown. Some quantities, marked with an asterisk (*), have been rescaled for comparison with other curves. (RIGHT) Pair gap as functions of $n$ for several interaction strengths $U$ (see color legend), with fixed $L=100$ and $k=0.002$.
\label{fig1}}
\end{figure}

In the context of the superfluid - insulator phase transition in harmonically confined chains \cite{sanino2024entanglement}, Fig. \ref{fig1} presents average quantities as a function of the system’s average density. All curves exhibit critical behavior near the transition point. Such non-monotonicity,
 is a well-known signature of quantum phase transitions. In confined systems, the emergence of a insulator suppresses the average occupancy of a single site $w_1$, which is intrinsically related to the entanglement of a single site $S$, as well as the charge fluctuations $\langle (\delta n)^2 \rangle$ and spin fluctuations $\langle (\delta s^z)^2 \rangle$, along with the superconducting order parameter $\Delta_{\mathrm{pair}}$. These average quantities serve as indicators of the transition, allowing the identification of a critical density $n_c$.

Focusing on the half-chain entanglement entropy $S_{L/2}$ in Fig. \ref{fig1}, we observe an abrupt drop to zero as soon as an insulating region emerges in the chain, even if it consists of a single insulating site. The critical densities $n_c$, defined as the values around which $S_{L/2}$ sharply vanishes, shift with increasing interaction strength. In the strong-coupling regime, pair formation is favored by attractive interactions, allowing full central occupation with fewer particles and thus lowering $n_c$. In contrast, under weak coupling, the transition shifts to higher densities, effectively extending the region of maximal $S_{L/2}$. Notably, when a insulator site forms at the center of the chain, the entanglement between the chain halves is abruptly suppressed, signaling the emergence of an insulating barrier between them.

In contrast to the half-chain entanglement, the presence of a quasi-metallic behavior between superfluid and insulator phases can be identified through the suppression of $\Delta_c$. Consistent with Sanino \textit{et al.} \cite{sanino2024entanglement}, the gap reaches small but finite values in the metallic regime at a critical density $n_c$ — corresponding to the minimum of $\Delta_c$ — which shifts depending on the $U/t$.

\subsection{Compressibility}

 
In addition, the charge gap for a pair of electrons~\cite{sanino2024entanglement} provides a measure of the energy required to insert or remove a pair of electrons from the chain, and it reads:
\begin{align}
    \Delta_{c} \equiv  E(N+2) + E(N-2) - 2E(N). 
\end{align}
Here, $E(N)$ is the ground-state energy with $N$ electrons.

We can notice that the charge gap is a finite-difference estimate of the second derivative of the energy with respect to the number of particles, $ \Delta_c \approx 4\frac{\partial^2 E(N)}{\partial N^2} = \frac{4}{\kappa L}$, where $\kappa$ is the compressibility. Since $\Delta_{c}$ is inversely proportional to the compressibility $\kappa$, a highly compressible system exhibits a small charge gap. In the BEC regime, the compressibility is large because electrons form tightly bound bosonic pairs that remain close to each other. As a consequence, adding or removing a coherent pair without breaking it becomes energetically favorable, resulting in a small value of $\Delta_c$. In contrast, in the BCS regime, the average pair separation is larger and the compressibility is reduced, leading to a larger charge gap, as we can clearly observe from Fig. \ref{fig_S}.

We can observe the SF–INS transition via $\Delta_c$ through its rapid increase, analogous to a low-compressibility regime. It is notable that, for densities slightly below the critical density for the SF–INS transition, the charge gap drops considerably and attains small but finite values; moreover, the position of this minimum shifts with $|U|$. The authors in Ref. \cite{sanino2024entanglement} termed this behavior a “metallic regime”.

Instead, our interpretation here is subtly different. First, it is important to recall that an effective tight-binding chain for BEC particles provides an accurate description of the overall behavior of the system in the BEC regime, particularly when $|U| \ge 6t$. As observed in Fig.~\ref{TTB_Spectrum} for $U=0$, the spectrum very close to the localized states exhibits a reduction in curvature, $\partial \epsilon_l / \partial l$, with $\partial^2 E / \partial N^2 \propto \partial \epsilon_l / \partial l$. In this regime, a larger number of particles can occupy nearby energy levels, leading to an enhancement of the compressibility $\kappa$, which in turn results in a smaller charge gap $\Delta_c$. Therefore, this minimum gap can be associated with the filling of states close to the localized region, where the compressibility is high, giving rise to a metallic-like behavior of the BEC pairs. Upon adding more particles to the chain, the system eventually undergoes a transition to the insulating phase. 

As discussed previously, increasing $|U|$ strengthens the effective confinement and binds electrons more tightly into pairs, such that fewer particles are required to reach the vicinity of the localized states. Consequently, the average density at which this minimum in the charge gap occurs is shifted toward lower values. This interpretation is also consistent with the results reported in Ref.~\cite{PhysRevB.82.014202}.

\section{Equivalent set of parameter}

\subsection{System size and density}

For fixed $k$, $U$, and $N$ in confined scenarios, once the confinement is strong enough that the density vanishes far from the trap center, increasing the system size does not change the physical properties of the system, provided the number of electrons $N=n.L$ is kept constant. Therefore, if
\begin{align}
    N = n.L = \mathrm{constant}(k,U),
\end{align}
the properties of the chain does not change, where $\mathrm{constant}(k,U)$ means an constant for fixed $k$ and $U$.

\subsection{Confinement strength and system size}

Let us notice that, if we take the limit $L \rightarrow \infty$ and define $j \rightarrow x$ in the continuum limit, with $c_{j\sigma} \rightarrow \Psi_\sigma(x)$, the Hamiltonian in Eq.~\eqref{Eq. 1} (with $j_0=0$) in the two body formalism can be written as:
\begin{align}
    H = \int_0^L dx  \sum_\sigma \Bigg[ t.kx^2  \Psi_\sigma^\dagger (x) \Psi_\sigma(x) -t \Psi_\sigma^\dagger (x) \big( \Psi_\sigma^\dagger (x-1) + \Psi_\sigma^\dagger (x+1) \big)  + \frac{U}{2}  \Psi_\sigma^\dagger (x) \Psi_{\bar\sigma}^\dagger(x)  \Psi_{\bar\sigma} (x)\Psi_\sigma(x) \Bigg].
\end{align}

By approximating the second derivative such as
\begin{align*}
    \partial_x^2  \Psi_\sigma (x) \approx   \Psi_\sigma (x+1) +   \Psi_\sigma (x-1) - 2 \Psi_\sigma (x),
\end{align*} we can rewrite this Hamiltonian as 
\begin{align}\label{Aux_E1}
    \frac{H}t = \int_0^L dx  \sum_\sigma  &\Psi_\sigma^\dagger (x) \Bigg[  kx^2 -\partial^2_x + \frac{U}{2t} \Psi_{\bar\sigma}^\dagger(x)  \Psi_{\bar\sigma} (x)\Bigg] \Psi_\sigma(x) .
\end{align}
Equation~\eqref{Aux_E1} contains the same information as Eq.~\eqref{Eq. 1}, except in the vicinity of a single site, where the approximation of the hopping term by second derivatives may fail.

Now, we can scale the Hamiltonian by the system size, and changing the variable by $\bar x = x/L$, resulting in
\begin{align}\label{Aux_E2}
    \frac{1}{t}\frac{H}L = \int_0^1 d \bar x  \sum_\sigma  &\Psi_\sigma^\dagger (\bar x ) \Bigg[  k L^2 {\bar x}^2 -\frac{\partial^2_{\bar x}}{L^2} + \frac{U}{2t} n_{\bar\sigma} (\bar x)\Bigg] \Psi_\sigma(\bar x).
\end{align}

\begin{figure}[tbh!]
\begin{centering}
\includegraphics[scale=0.65]{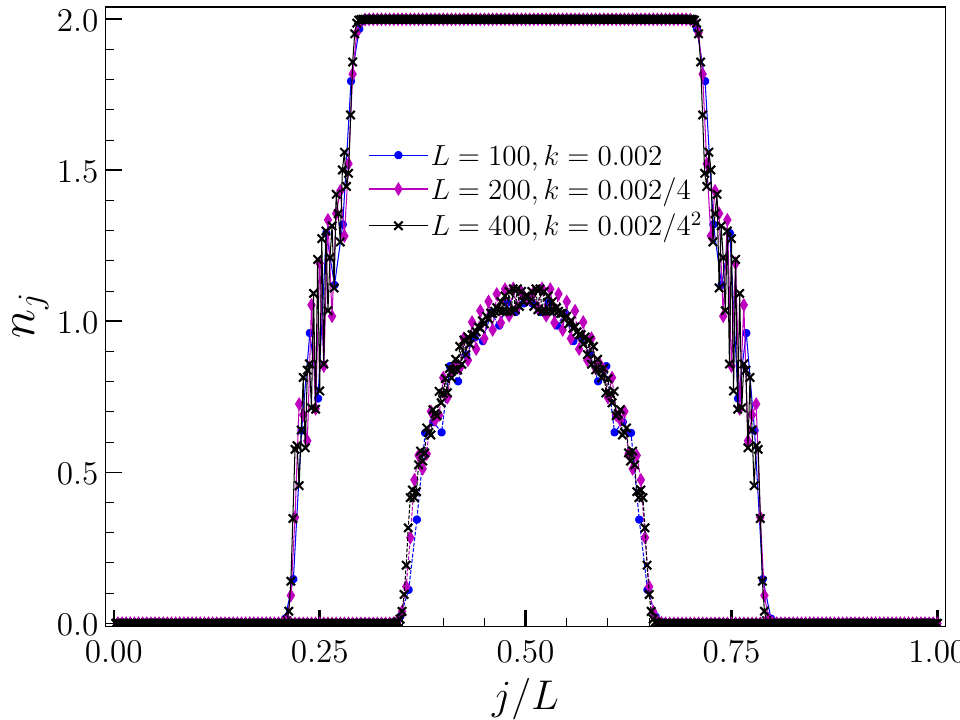}
\par\end{centering}
\caption{Results for the density profile along the chain as a function of $(j/L)$ for $n=1$ (larger effective chain) and $n=0.25$ (smaller effective chain), obtained for different values of $k$ and $L$ while keeping $kL^2$ constant for fixed $U=-10t$.
\label{scaling_kxL}}
\end{figure}

Therefore, in harmonically confined scenarios and for sufficiently large $L$, the derivative term of Eq.~\eqref{Aux_E2} contributes little at large spatial scales. The confinement and Coulomb terms then dominate, and the scaled density distribution as a function of $\bar x=j/L$ becomes approximately invariant if
\begin{align}
    kL^2 = \mathrm{constant}(U,n),
\end{align}
as shown in Fig.~\ref{scaling_kxL}. Specifically, Fig.~\ref{scaling_kxL} shows results for $n=1.0$ and $n=0.25$ obtained for different values of $k$ and $L$ satisfying constant $kL^2$. These results show that rescaling the system size yields nearly identical density profiles, except for small differences due to density oscillations, but the overall shape is kept constant.  
 
\end{document}